\documentstyle[epsf,preprint,eqsecnum,,aps]{revtex}
\begin{document}
\draft
\title{On Migdal's theorem and the pseudogap}

\author{P. Monthoux}
\address{Cavendish Laboratory, University of Cambridge
\\Madingley Road, Cambridge CB3 0HE, United Kingdom}
\date{\today}
\maketitle

\begin{abstract}

We study a model of quasiparticles on a two-dimensional
square lattice coupled to Gaussian distributed dynamical
molecular fields. We consider two types of such fields,
a vector molecular field that couples to the quasiparticle 
spin-density and a scalar field coupled to the quasiparticle
number density. The model describes quasiparticles coupled 
to spin or charge fluctuations, and is solved by a Monte
Carlo sampling of the molecular field distributions.
When the molecular field correlations are sufficiently
weak, the corrections to the self-consistent
Eliashberg theory do not bring about qualitative changes
in the quasiparticle spectrum. But for a range of model 
parameters near the magnetic boundary, we find that 
Migdal's theorem does not apply and the quasiparticle 
spectrum is qualitatively different from its Eliashberg 
approximation. In the range of model parameters studied, 
we find the transverse spin-fluctuation modes 
play a key role. While a pseudogap opens when 
quasiparticles are coupled to antiferromagnetic 
fluctuations, such a pseudogap is not observed in the 
corresponding charge-fluctuation case for the range
of parameters studied, where vertex corrections are found 
to effectively reduce the strength of the interaction. 
This suggests that one has to be closer to the border
of long-range order to observe pseudogap effects in the
charge-fluctuation case than for a spin-fluctuation 
induced interaction under otherwise similar conditions.

An important feature of the magnetic
pseudogap found in the present calculations is that it
is strongly anisotropic. It vanishes along the diagonal
of the Brillouin zone and is large near the zone boundary.
In the case of ferromagnetic fluctuations, we also find a 
range of model parameters with qualitative changes in 
the quasiparticle spectral function not captured by 
the one-loop approximation, in that the quasiparticle 
peak splits into two. We find that one needs to be closer
to the magnetic boundary to observe the pseudogap effects
in the nearly ferromagnetic case relative to the nearly
antiferromagnetic one, under otherwise similar conditions.
We provide intuitive arguments to explain the physical 
origin of the breakdown of Migdal's theorem.

\end{abstract}

\pacs{PACS Nos. 71.27.+a}

\narrowtext

\section{Introduction}

The polarizer-analyser analogy provides an intuitive 
description of the effective interaction between 
quasiparticles in a quantum many-body system. The first 
quasiparticle polarizes the medium in which it travels 
and the second quasiparticle, the analyser, feels 
the disturbance induced by the first one. 
In a strongly correlated system one can expect this 
induced polarization of the medium to be very complex, 
and in practice simplifying assumptions are made. A 
commonly used approximation is the description of 
the polarization effects by the appropriate
linear response function of the material. 
Furthermore, one typically only considers 
the interaction channel for which the linear
response of the system is the largest. On the border of 
long-range magnetic order for instance, the spin-spin
correlation function is the most enhanced and it is 
plausible that the dominant interaction channel is of 
magnetic origin and depends on the relative spin 
orientations of the interacting quasiparticles.

It has been shown that such a magnetic interaction,
treated in the self-consistent Eliashberg approximation, 
can produce anomalous normal state properties and 
superconducting instabilities to anisotropic pairing states.  
It correctly predicted the symmetry of the Cooper state in the 
copper oxide superconductors\cite{DwavePrediction} and is 
consistent with spin-triplet p-wave pairing in superfluid 
$^3He$ [for a recent review see, e.g., Ref.~\cite{Dobbs}].  
One also gets the correct order of magnitude of the 
superconducting and superfluid transition temperature 
$T_c$ when the model parameters are inferred from 
experiments in the normal state of the above systems.
For the case of a nearly half-filled single band, 
the calculations showed that the Eliashberg superconducting 
transition temperature $T_c$ is higher for the tetragonal 
(quasi-2D) than cubic (3D) lattice\cite{Nakamura,Arita,ML2,ML3}. 
Particularly striking is the comparison between the cubic 
antiferromagnetic metal $CeIn_3$ and the closely related
compound $CeCoIn_5$\cite{Fisk}. Superconductivity is found 
to extend over a much wider range in both temperature and 
pressure in $CeCoIn_5$ than in $CeIn_3$. These findings and 
the growing evidence that the pairing symmetry in $CeCoIn_5$
is d-wave in character, were correctly anticipated by the
magnetic interaction model.

While there have been a number of examples of 
superconductivity on the border of antiferromagnetism,
the corresponding phenomenon on the edge of metallic 
ferromagnetism has only been found recently. This result
is not surprising within the framework of the magnetic 
interaction model. For otherwise equivalent conditions,
the superconducting transition temperature is typically
much higher on the border of antiferromagnetism than on
the border of ferromagnetism\cite{ML2,ML3,ML1}. An
intuitive understanding of this finding pointed to
candidate systems in which superconductivity on the
border of ferromagnetism would be more likely to be
observed. In particular, one ought to look for systems
with strong spin anisotropy, i.e, with stong spin-orbit
coupling and/or in a weakly spin-polarized state. This
suggested a more detailed investigation of $UGe_2$ at
high pressure which satisfied the above conditions and
could be prepared in high purity form. This material
proved to be the first example of the co-existence of
superconductivity and itinerant-electron 
ferromagnetism\cite{Saxena}.

The self-consistent Eliashberg treatment of the magnetic 
interaction model can produce strongly damped 
quasiparticles, but the electronic spectral function
one obtains always shows a quasiparticle peak as
one approaches the Fermi level. This is in qualitative 
disagreement with photoemission experiments on underdoped 
cuprate superconductors which show a near absence of a 
quasiparticle peak near the $(\pi,0)$ point in the
Brillouin zone\cite{Photoemission}. This depletion of 
quasiparticle states, or pseudogap, is also seen in 
thermodynamic measurements\cite{SpecificHeat}. The
phenomenon may not be specific to the underdoped cuprates.
For instance, the possible existence of a pseudogap 
in the heavy fermion compound $CeCoIn_5$ has recently 
been reported\cite{115}.

Is the disagreement between theory and experiment the 
manifestation of a fundamental flaw in the approach or 
does it simply reflect the inadequacy of the approximations 
used in the solution of the model? Questions regarding the 
validity of the Eliashberg treatment of the magnetic interaction
model have been raised\cite{Schrieffer}. The one-loop 
approximation effectively assumes that quasiparticles 
behave as test particles. There must therefore be 
corrections to the simple theory, referred to as vertex 
corrections, coming from the fact that real and test 
particles behave differently. One can expect
these vertex corrections to produce quantitative 
changes to the self-consistent Eliashberg theory as 
one approaches the border of long-range magnetic order. 
In Ref.~\cite{Vertex} it was shown that for optimally doped 
cuprates, vertex corrections did not bring about significant 
changes to the single spin-fluctuation approximation, 
producing an enhancement of the spin-fluctuation 
interaction of the order of 20\%. The sign of this 
correction\cite{Vertex} is opposite to that expected 
in the case of a phonon-mediated interaction~\cite{Doug} 
and in the spin-density-wave phase\cite{Schrieffer}. 
This enhancement of the fermion spin-fluctuation vertex 
in the paramagnetic state is due to the transverse 
spin-fluctuation modes.

It has been argued, in particular by 
Schrieffer\cite{Schrieffer}, that as the 
antiferromagnetic correlations got stronger, 
the system should display characteristics akin 
to the antiferromagnetic insulating state and 
that the behavior of the quasiparticles should 
become qualitatively different from that of a simple 
metal. The results presented here show that this 
physical insight is correct in that Midgal's theorem can 
qualitatively break down when the antiferromagnetic 
correlations become strong enough and a different state 
emerges. 

The validity of Baym-Kadanoff many-body theories such 
as the fluctuation exchange approximation\cite{FLEX}, 
which is in many ways similar to the Eliashberg theory 
of the magnetic interaction model, has been extensively 
studied by Tremblay and collaborators in the context of 
the Hubbard model\cite{Vilk1,Vilk2,Moukouri}. They find that
close to the magnetic boundary, Migdal's theorem
qualitatively breaks down, in that a critical-fluctuation-induced 
pseudogap (or precursor pseudogap) is observed in the
Quantum Monte Carlo simulations but is not found in
the fluctuation exchange approximation. 
Similar results about the role of vertex corrections
were reported in the case of the attractive Hubbard 
model\cite{Vilk3,Allen1,Allen2,Kyung,Fujimoto}, where 
in this case the precursor pseudogap is caused by
critical pairing fluctuations.

In this paper, we examine corrections to the single 
spin-fluctuation exchange approximation in two dimensions
using a non-perturbative formulation of the magnetic 
interaction model which is amenable to computer 
simulation\cite{PhilMag}. We find that as the 
antiferromagnetic correlations become 
strong enough for vertex corrections 
to produce a qualitative change relative to the one-loop 
approximation, a pseudogap opens in the quasiparticle 
spectrum. In this respect, our results are similar 
to those obtained for the Hubbard 
model\cite{Vilk1,Vilk2,Moukouri}.
The pseudogap we find is strongly anisotropic in that
it vanishes along the diagonal of the Brillouin zone and
is large near the zone boundary. We demonstrate that, 
in the range of model parameters studied here, the 
transverse spin-fluctuation modes are key to the 
appearance of the pseudogap by considering the case of 
commensurate charge-fluctuations with a spectrum identical 
to that of the paramagnons. While a pseudogap opens when 
quasiparticles are coupled to magnetic fluctuations, 
such a pseudogap is not observed in the 
corresponding charge-fluctuation case for the range
of parameters studied, where vertex corrections are found 
to effectively reduce the strength of the interaction. 
This suggests that one has to be closer to the border
of long-range order to observe pseudogap effects in the
charge-fluctuation case than for a spin-fluctuation 
induced interaction under otherwise similar conditions.
In the case of nearly ferromagnetic systems, as magnetic
correlations get stronger, we also find qualitative 
changes in the quasiparticle spectral function not 
captured by the one-loop approximation. The quasiparticle 
peak splits into two distinct peaks with a lowering 
of the tunneling density of states at the Fermi
level. For the range of parameters studied, this 
suppression of the tunneling density of states 
at the Fermi level is weaker than in the case of nearly 
antiferromagnetic systems for otherwise similar magnetic 
and electronic spectrum parameters.

The paper is organized as follows. In the next section 
we describe the model and the class of vertex corrections 
considered. Section III contains the results of the 
numerical simulations. In section IV, we give intuitive
arguments for the physical origin of the pseudogap. 
Finally we give a summary and outlook. Most of the 
technical details are included in an appendix.

\section{Model}

There are essentially two different ways to describe quantum 
many-body systems. In the Newtonian or Hamiltonian approach,
one considers particles interacting with each other via
pairwise interaction potentials. This is the point of view
commonly adopted in the perturbation-theoretic approach to
the non-relativistic many-electron problem. In the relativistic 
version of the theory, however, one adopts a Maxwellian point 
of view in which the interactions between electrons are mediated 
by a field, the quantized electromagnetic field. The Maxwellian
approach is also widely used to carry out numerical simulations
of interacting systems based on the Feynman path integral. 
In such functional integrals, bosonic fields are represented by
c-numbers. But fermion fields must be represented by 
anticommuting, or Grassmann, numbers that are not easily 
handled by digital computers. When a dynamical molecular 
(or Hubbard-Stratonovich) field is introduced to mediate the 
interactions between the fermions, the problem is reduced 
to that of non-interacting particles in a fluctuating field. 
The integrals over the anticommuting variables can then 
be evaluated exactly, at least formally, thus eliminating 
the troublesome Grassmann numbers from the problem. In the 
following, we adopt the Maxwellian point of view and 
use dynamical molecular fields to mediate the interactions
between quasiparticles.

In general, the distribution of Hubbard-Stratonovich fields
is very complex and usually leads to intractable problems
due to the infamous "fermion sign problem". In a many-body
system, the presence of other particles produces changes 
in the effective interaction between two particles through
screening effects and induces arbitrarily complex 
self-interactions of the dynamical molecular fields.

In this paper, we assume that the renormalization of the 
effective two-body interaction can be accounted for by
a redefinition of the parameters entering the
bare interaction. And we ignore all the self-interactions
of the Hubbard-Stratonovich fields. The model was 
introduced in Ref.~\cite{PhilMag} and bears some resemblance
to the "quenched approximation" of lattice gauge theories 
introduced by Marinari et al., who, incidentally, chose the 
name by analogy to condensed matter physics\cite{quenched}.
A recent application of the quenched approximation to 
the pseudogap problem for static Hubbard-Stratonovich
fields and a clear exposition of the formalism
is given by Posazhennikova and Coleman\cite{Coleman}.
But in the present work, however, we use dynamical rather 
than static Hubbard-Stratonovich fields and a non-separable 
form of the molecular field correlation function.

Very close to the boundary of magnetic or charge 
long-range order, the self-interactions of 
the dynamical molecular field ignored in the 
present work are known to be important for low dimensional 
systems\cite{HertzMillis}. The approximations made here may
not be appropriate in a number of other cases. The virtue of
the present approach is that it gives insights into quantum 
many-body problems that are essentially non-perturbative.
More importantly, the results presented in this paper 
demonstrates that the simplest model already yields 
interesting physics. Some of the simplifications 
made here can in principle be relaxed and the theory
extended accordingly.

To be more specific, we consider particles on a two-dimensional
square lattice whose Hamiltonian in the absence of interactions
is

\begin{equation}
{\hat h}_0(\tau) = -\sum_{i,j,\alpha} 
t_{ij}\psi^\dagger_{i\alpha}(\tau)\psi_{j\alpha}(\tau)
- \mu \sum_{i\alpha}\psi^\dagger_{i\alpha}(\tau)\psi_{i\alpha}(\tau)
\label{ham0}
\end{equation}

\noindent where $t_{ij}$ is the tight-binding hopping matrix,
$\mu$ the chemical potential and $\psi^\dagger_{i\alpha}$,
$\psi_{i\alpha}$ respectively create and annihilate a fermion 
of spin orientation $\alpha$ at site $i$. We take $t_{ij} = t$ 
if sites $i$ and $j$ are nearest-neighbors and $t_{ij} = t'$ 
if sites $i$ and $j$ are next-nearest-neighbors. 

To introduce interactions between the particles, we couple 
them to a dynamical molecular (or Hubbard-Stratonovich) 
field. It is instructive to consider two different types
of molecular fields. In the first instance, we consider a
vector Hubbard-Stratonovich field that couples locally to
the fermion spin density. This is the case considered in 
Ref.~\cite{PhilMag} and describes quasiparticles coupled to
magnetic fluctuations. To illustrate the role played by 
transverse spin-fluctuations we shall also consider the case
of a scalar field that couples locally to the fermion
number density. This case corresponds to a coupling to
charge-fluctuations or, within the approximation we are using 
here, "Ising"-like magnetic fluctuations where only longitudinal 
modes are present. The Hamiltonians at imaginary time $\tau$ 
for particles coupled to the fluctuating exchange or scalar 
dynamical field are then

\begin{eqnarray}
{\hat h}(\tau) & = & {\hat h}_0(\tau) - {g\over \sqrt{3}}\sum_{i\alpha\gamma}
{\bf M}_i(\tau)\cdot\psi^\dagger_{i\alpha}(\tau){\bf \sigma}_{\alpha\gamma}
\psi_{i\gamma}(\tau) 
\label{ham1} \\
{\hat h}(\tau) & = & {\hat h}_0(\tau) - g\sum_{i\alpha}
\Phi_i(\tau)\psi^\dagger_{i\alpha}(\tau)\psi_{i\alpha}(\tau)
\label{ham2}
\end{eqnarray}

\noindent where ${\bf M}_i(\tau) = (M^x_i(\tau),M^y_i(\tau),M^z_i(\tau))^T$
and $\Phi_i(\tau)$ are the real vector exchange and scalar 
Hubbard-Stratonovich fields respectively, and $g$ the 
coupling constant. The reason for the choice of an extra 
factor $1/\sqrt{3}$ in Eq.~(\ref{ham1}) will become clear later.

Since we ignore the self-interactions of the molecular 
fields, their distribution is Gaussian and given by\cite{PhilMag}

\begin{eqnarray}
{\cal P}[{\bf M}] & = & {1\over Z} \exp\Bigg(-\sum_{{\bf q},\nu_n} 
{{\bf M}({\bf q},i\nu_n)\cdot {\bf M}(-{\bf q},-i\nu_n)
\over 2\alpha({\bf q},i\nu_n)}\Bigg)
\label{ProbM} \\
Z & = & \int D{\bf M}\exp\Bigg(-\sum_{{\bf q},\nu_n} 
{{\bf M}({\bf q},i\nu_n)\cdot {\bf M}(-{\bf q},-i\nu_n)
\over 2\alpha({\bf q},i\nu_n)}\Bigg)
\label{NormM}
\end{eqnarray}

\noindent in the case of a vector exchange molecular field and

\begin{eqnarray}
{\cal P}[\Phi] & = & {1\over Z}\exp\Bigg(-\sum_{{\bf q},\nu_n} 
{\Phi({\bf q},i\nu_n)\Phi(-{\bf q},-i\nu_n)
\over 2\alpha({\bf q},i\nu_n)}\Bigg)
\label{ProbPhi} \\
Z & = & \int D\Phi\exp\Bigg(-\sum_{{\bf q},\nu_n} 
{\Phi({\bf q},i\nu_n)\Phi(-{\bf q},-i\nu_n)
\over 2\alpha({\bf q},i\nu_n)}\Bigg)
\label{NormPhi}
\end{eqnarray}

\noindent in the case of a scalar Hubbard-Stratonovich field. 
In both cases $\nu_n = 2\pi n T$ since the dynamical molecular
fields are periodic functions in the interval $[0,\beta=1/T]$.
The Fourier transforms of the molecular fields are defined as

\begin{eqnarray}
{\bf M}_{\bf R}(\tau) & = & \sum_{{\bf q},\nu_n} {\bf M}({\bf q},i\nu_n) 
\exp\Big(-i[{\bf q}\cdot{\bf R}-\nu_n\tau]\Big)
\label{FourierM} \\
\Phi_{\bf R}(\tau) & = & \sum_{{\bf q},\nu_n} \Phi({\bf q},i\nu_n)
\exp\Big(-i[{\bf q}\cdot{\bf R}-\nu_n\tau]\Big)
\label{FourierP}
\end{eqnarray}

We are considering the case where there is no long-range 
magnetic or charge order. The average of the dynamical
molecular fields must then vanish and their Gaussian
distributions Eqs.~(\ref{ProbM},\ref{ProbPhi}) are
completely determined by their variance 
$\alpha({\bf q},i\nu_n)$, which we take to be

\begin{equation}
\alpha({\bf q},i\nu_n) = \cases{{1\over 2}{T\over N}\chi({\bf q},i\nu_n)
&if ${\bf M}({\bf q},i\nu_n)$ or $\Phi({\bf q},i\nu_n)$ complex \cr
{T\over N}\chi({\bf q},i\nu_n)
&if ${\bf M}({\bf q},i\nu_n)$ or $\Phi({\bf q},i\nu_n)$ real}
\label{alpha}
\end{equation}

\noindent where $N$ is the number of allowed wavevectors
in the Brillouin zone. Then

\begin{eqnarray}
\Big<M_i({\bf q},i\nu_n)M_j({\bf k},i\Omega_n)\Big> & = &
{T\over N}\chi({\bf q},i\nu_n)\delta_{{\bf q},-{\bf k}}
\delta_{\nu_n,-\Omega_n}\delta_{i,j}
\label{Mcorr} \\
\Big<\Phi({\bf q},i\nu_n)\Phi({\bf k},i\Omega_n)\Big> & = &
{T\over N}\chi({\bf q},i\nu_n)\delta_{{\bf q},-{\bf k}}
\delta_{\nu_n,-\Omega_n}
\label{Pcorr}
\end{eqnarray}

\noindent where $<\dots>$ denotes an average over the 
probability distributions Eq.~(\ref{ProbM}) and 
Eq.~(\ref{ProbPhi}) for the vector and scalar cases
respectively. In order to compare the scalar and vector 
molecular fields, we take the same form for their
correlation function $\chi({\bf q},i\nu_n)$
and parametrize it as in Refs.~\cite{ML2,ML1}. 
In what follows, we set the lattice spacing $a$ to unity.
For real frequencies, we have

\begin{equation}
\chi({\bf q},\omega) = {\chi_0\kappa_0^2\over \kappa^2 + \widehat{q}^2 
- i{\omega\over \eta(\widehat{q})}}
\label{chiML}
\end{equation}

\noindent where $\kappa$ and $\kappa_0$ are the correlation 
wavevectors or inverse correlation lengths in units of 
the lattice spacing, with and without strong 
correlations, respectively.  Let

\begin{equation}
\widehat{q}_{\pm}^2 = 4 \pm 2(\cos(q_x)+\cos(q_y)) 
\label{qdef}
\end{equation}

We shall consider commensuate charge-fluctuations and
antiferromagnetic spin-fluctuations, in which case
the parameters $\widehat{q}^2$ and $\eta(\widehat{q})$ 
in Eq.~(\ref{chiML}) are defined as

\begin{eqnarray}
\widehat{q}^2 & = & \widehat{q}_{+}^2 \\
\eta(\widehat{q}) & = & T_0\widehat{q}_{-}
\label{antiferro}
\end{eqnarray}

\noindent where $T_0$ is a characteristic temperature.  

We shall also consider the case of ferromagnetic 
spin-fluctuations, where the parameters 
$\widehat{q}^2$ and $\eta(\widehat{q})$ 
in Eq.~(\ref{chiML}) are given by

\begin{eqnarray}
\widehat{q}^2 & = & \widehat{q}_{-}^2 \\
\eta(\widehat{q}) & = & T_0\widehat{q}_{-}
\label{ferro}
\end{eqnarray}

\noindent $\chi({\bf q},i\nu_n)$ is related to the imaginary 
part of the response function $Im\chi({\bf q},\omega)$, 
Eq.~(\ref{chiML}), via the spectral representation

\begin{equation}
\chi({\bf q},i\nu_n) = -\int_{-\infty}^{+\infty}{d\omega\over \pi}
{Im\chi({\bf q},\omega)\over i\nu_n - \omega}
\label{chi_mats}
\end{equation}

\noindent To get $\chi({\bf q},i\nu_n)$ to decay as $1/\nu_n^2$ as
$\nu_n \rightarrow \infty$, as it should, we introduce a cutoff 
$\omega_0$ and take $Im\chi({\bf q},\omega) = 0$ for $\omega 
\geq \omega_0$. A natural choice for the cutoff is $\omega_0 
= \eta(\widehat{q})\kappa_0^2$.

In the approximation we are considering, the single 
particle Green's function is the average over the probability
distributions ${\cal P}[{\bf M}]$ (Eq.~(\ref{ProbM})) or 
${\cal P}[\Phi]$ (Eq.~(\ref{ProbPhi})) of the fermion
Green's function in a dynamical vector or scalar field.

\begin{eqnarray}
{\cal G}(i\sigma\tau ; j\sigma'\tau') & = & 
\int D{\bf M}\; {\cal P}[{\bf M}]\; G(i\sigma\tau ; j\sigma'\tau'|[{\bf M}])
\label{gM} \\
{\cal G}(i\sigma\tau ; j\sigma'\tau') & = & 
\int D\Phi\; {\cal P}[\Phi]\; G(i\sigma\tau ; j\sigma'\tau'|[\Phi])
\label{gPhi}
\end{eqnarray}

\noindent where

\begin{equation}
G(i\sigma\tau ; j\sigma'\tau'|[{\bf M}]\;or \;[\Phi]) 
= -\big<T_\tau\{\psi_{i\sigma}(\tau)
\psi^\dagger_{j\sigma'}(\tau')\}\big> 
\label{gfield} 
\end{equation}

\noindent is the single particle Green's function in a
dynamical molecular field and is discussed in the appendix.
In evaluating expressions Eqs.~(\ref{gM},\ref{gPhi}) one
is summing over all Feynman diagrams corresponding to
spin or charge-fluctuation exchanges\cite{PhilMag,Coleman}.
The diagrammatic expansion of the Green's function 
in a dynamical field, Eq.~(\ref{gfield}) and its 
average, Eq.~(\ref{gM}) are shown pictorially in Fig.~1.
A non-perturbative study of quasiparticles coupled
to static magnetic fluctuations using a diagram
summation technique was reported in Ref.~\cite{Static}.
In this paper we do not resort to a diagrammatic
expansion but rather evaluate the averages in 
Eqs.~(\ref{gM},\ref{gPhi}) over the dynamical
molecular fields by Monte Carlo sampling.

It is very instructive to compare the results of
the Monte Carlo simulations with the self-consistent
Eliashberg calculations for the same model. If one
only considers single spin or charge-fluctuation
exchange processes, the single particle Green's
function is given by:

\begin{equation}
\Sigma({\bf p},i\omega_n) = g^2{T\over N}\sum_{\Omega_n}
\sum_{\bf k}\chi({\bf p}-{\bf k},i\omega_n-i\Omega_n)
{\cal G}({\bf k},i\Omega_n)
\label{SigmaE}
\end{equation}

\begin{equation}
{\cal G}({\bf p},i\omega_n) = {1\over i\omega_n 
- (\epsilon_{\bf p}-\mu) - \Sigma({\bf p} ,i\omega_n)}
\label{GreenE}
\end{equation}

\noindent where $\Sigma({\bf p},i\omega_n)$ is the 
quasiparticle self-energy, ${\cal G}({\bf p},i\omega_n)$ 
the one-particle Green's function. $\epsilon_{\bf p}$ 
is the tight-binding dispersion relation obtained from 
Fourier transforming the hopping matrix $t_{ij}$ in 
Eq.~(\ref{ham0}) and $\mu$ the chemical potential.
The choice of the factor $1/\sqrt{3}$ in Eq.~(\ref{ham1})
means one obtains the same one-loop equations in the
the case of an exchange molecular field ${\bf M}$
and that of the scalar field $\Phi$, thereby
simplifying the comparison between the two cases.

\section{Results}

The quasiparticle dispersion relation for the 
two-dimensional square lattice is obtained from 
Eq.~(\ref{ham0}). We measure all energies and temperatures 
in units of the nearest-neighbor hopping parameter $t$. 
We set the next-nearest-neighbor hopping parameter 
$t'=-0.45t$. The chemical potential is adjusted so that
the electronic band filling is $n=0.9$. The dimensionless 
parameters describing the molecular field correlations
are $g^2\chi_0/t$, $T_0/t$, $\kappa_0$ and $\kappa$. 
A complete exploration of the parameter space of the 
model is beyond the scope of this preliminary study.
We chose a representative value for $\kappa_0^2 = 12$, 
and set $T_0 = 0.67t$ as in our earlier work\cite{ML2,ML1}. 
For an electronic bandwidth of $1eV$, $T_0\approx 1000^\circ$K.
We only consider one value of the coupling constant
$g^2\chi_0/t = 2$. In the random phase approximation,
the magnetic instability would be obtained for a value
of $g^2\chi_0/t$ of the order of 10. We consider what 
happens to the quasiparticle spectrum at a fixed 
temperature $T = 0.25t$ as the inverse correlation 
length $\kappa$ changes. 

The calculations were done on a 16 by 16 lattice,
with 41 imaginary time slices, or equivalently 41
Matsubara frequencies for the molecular fields, 
${\bf M}({\bf q},i\nu_n)$ and $\Phi({\bf q},i\nu_n)$
($\nu_n = 2\pi n T$, with $ n=0,\pm 1,\dots,\pm 20$).

By analytic continuation of the single particle Green's 
function ${\cal G}({\bf k},\tau)$ or 
${\cal G}({\bf k},i\omega_n)$ one can obtain the 
quasiparticle spectral function $A({\bf k},\omega) 
= - {1\over \pi}Im\;{\cal G}_R({\bf k},\omega)$ and 
the tunneling density of states $N(\omega) = {1\over N}
\sum_{\bf k}A({\bf k},\omega)$, where
${\cal G}_R({\bf k},\omega)$ is the retarded single
particle Green's function. In the case of the one-loop
approximation, ${\cal G}({\bf k},i\omega_n)$ is 
analytically continued from imaginary to real frequencies 
by means of Pad\'e approximants\cite{Serene}. The 
imaginary time Monte Carlo data is analytically 
continued with the Maximum Entropy method\cite{MaxEnt}.
We have used 2000 MC samples binned in groups of 20
to make 100 measurements. We have used three different
versions of the Maximum Entropy method. The Classic 
MaxEnt and two versions of Average MaxEnt, which is the
method recommended in Ref.~\cite{MaxEnt}, where the
probability for the $\alpha$ parameter is  either a constant
or proportional to $1/\alpha$\cite{MaxEnt}. In all cases
we chose a flat default model. The results for the three 
different versions of the Maximum Entropy method tried 
were nearly identical in the case of the vector molecular
field. There were slight differences between the Classic
and Average MaxEnt solutions for the spectral function in
the scalar molecular field case, while the two versions 
of the Average MaxEnt gave nearly identical results. The 
choice of the $\sim 1/\alpha$ probability distribution
in the Average MaxEnt method gave a slightly better 
fit to the MC data and all the results shown here 
are those obtained with this choice of probability 
distribution for the $\alpha$ parameter.

Fig.~2 shows the tunneling density of states $N(\omega)$ and
the spectral function $A({\bf k},\omega) = 
-{1\over\pi}Im{\cal G}_R({\bf k},\omega)$ for the one-loop 
Eliashberg approximation to the Green's function, 
Eqs.~(\ref{SigmaE},\ref{GreenE}), in the case of
commensurate charge-fluctuations and antiferromagnetic
spin-fluctuations. Figs.~2b,c show a strong anisotropy
of the spectral function. $A({\bf k},\omega)$ is sharper
when ${\bf k}$ is along the diagonal compared to the case
where  ${\bf k}$ is near a hot spot (which is a point 
on the Fermi surface accessible from another via a 
momentum transfer of ${\bf Q} = (\pi,\pi)$). Also note 
the monotonic broadening of $A({\bf k},\omega)$ as 
the parameter $\kappa^2$ is reduced.

In Fig.~3 we show the tunneling density of states and 
quasiparticle spectral function one obtains from the
Monte Carlo calculation with a coupling to the scalar
dynamical molecular field $\Phi$. By comparison to the 
one-loop self-consistent results of Fig.~2, the vertex 
corrections give rise to a sharpening of the quasiparticle 
spectral function, except for $\kappa^2=4$ and 
${\bf k} = (3\pi/8,3\pi/8)$. For the range of values 
of model parameters considered here, the multiple 
charge-fluctuation exchanges do not lead to a breakdown 
of the quasiparticle picture. Note that contrary
to the Eliashberg result, as $\kappa^2$ is reduced
from 4, the spectral function initially sharpens before
broadening again at the lower values of $\kappa^2$ 
considered here. For $\kappa^2=4,2$, the spectral function
is sharper near the hot spot than along the diagonal, 
again in contrast to the one-loop self-consistent result. 
It is approximately isotropic at $\kappa^2 = 1$ and becomes
sharper along the diagonal than near the hot spot at
the lower values of $\kappa^2=0.50,0.25$. Quite 
generally Fig.~3 shows that when vertex corrections 
are included the spectral function is less anisotropic
than the Eliashberg result for the values of $\kappa^2$
considered.

The Monte Carlo results for quasiparticles coupled to
antiferromagnetic spin-fluctuations are shown in Fig.~4.
For $\kappa^2 \leq 1$, a pseudogap appears at the point
${\bf k} = (\pi,\pi/8)$ but not along the diagonal, in
qualitative agreement with experiments on underdoped
cuprates\cite{Photoemission}. The pseudogap 
also shows up in the tunneling density of 
states $N(\omega)$ which is suppressed at 
the Fermi level for $\kappa^2 \leq 1$. By comparing
the spectral function of Figs.~4b,c to the one-loop 
self-consistent result, we see that for the values of 
$\kappa^2$ where there isn't a pseudogap, vertex corrections 
lead to a broadening of the spectral function and a reduction
of the momentum anisotropy of the spectral function. 
The contrast between the non-perturbative calculations 
for charge and spin-fluctuations and the comparison 
with the Eliashberg result is shown in Fig.~5 for 
$\kappa^2=2$. In this case it is clear that the effect 
of vertex corrections is to make the spectral function 
sharper than its Eliashberg approximation in the 
charge-fluctuation case and broader when the quasiparticles 
are coupled to magnetic fluctuations. For these model 
parameters, corrections to the Eliashberg approximation 
reduce the effective charge-mediated interaction and 
enhance the coupling to magnetic fluctuations.

In Fig.~6 the results of the one-loop calculations in
the case of ferromagnetic spin-fluctuation exchange
are shown. As could be expected, the spectral function 
for quasiparticles near the Fermi level does not depend 
strongly on whether the momentum is near the hot spot, 
Fig.~6b, or along the diagonal in the Brillouin zone, 
Fig.~6c, in contrast to the nearly antiferromagnetic 
case. At the Eliashberg level, the spectral function
broadens monotonically as the correlation wavevector
$\kappa$ is reduced and is broader than the 
corresponding nearly antiferromagnetic case, 
Fig.~2b,c.

The Monte Carlo results for quasiparticles coupled to
ferromagnetic spin-fluctuations are shown in Fig.~7.
The tunneling density of states at the Fermi level
$N(\omega=0)$ begins to drop as the parameter 
$\kappa^2 \leq 0.50$. For $\kappa^2 = 0.25$, the 
quasiparticle peak has effectively split into two
peaks, and a precursor of this effect can be seen
at $\kappa^2 = 0.50$. This phenomenon is qualitatively
different from what one can obtain at the one-loop 
level, Fig.~6, just as the pseudogap seen in Fig.~4,
when the quasiparticle interactions are mediated
by an antiferromagnetically correlated exchange field. 
For the larger value of $\kappa^2$ where the splitting 
of the quasiparticle peak is not observed, a comparison
of the Monte Carlo results with the one-loop self-consistent
calculations shown in Fig.~6 shows that vertex corrections
bring about a broadening of the quasiparticle
spectral function. Therefore, just as in the 
antiferromagnetic case, vertex corrections enhance 
the magnetically-mediated interaction, albeit to a 
lesser degree, as a close look at Figs.~2,4 and
6,7 indicates. This could explain why qualitative 
changes from the Eliashberg solution are seen for smaller 
values of $\kappa^2$ for coupling to ferromagnetic than 
antiferromagnetic fluctuations.

\section{Discussion}

For the model considered here, at the one-loop level 
the exchange of commensurate charge-fluctuations 
and antiferromagnetic spin-fluctuations yield the 
same quasiparticle properties. Once vertex
corrections are included, the differences between 
the two cases are evident. In all but one case, 
${\bf k} = (3\pi/8,3\pi/8)$ and $\kappa^2 = 4$,  
corrections to the one-loop approximation make the 
quasiparticle peak sharper than the Eliashberg result for 
charge-fluctuations. In the case of antiferromagnetic 
spin-fluctuations, the quasiparticle peak gets broader
as vertex corrections are included and a pseudogap
appears for the lower values of $\kappa^2$ studied.
The difference between the scalar and vector molecular
fields appears at the leading vertex correction 
to the one-loop approximation\cite{Vertex} 
shown in Fig.~8. The frequency and momentum integrals
are the same in both cases since we assume the commensurate
charge-fluctuations and antiferromagnetic spin-fluctuations
have the same spectrum. However, the spin sums in the two
cases are not identical. In the case of a coupling of the 
molecular field to the quasiparticle spin-density,
one gets a factor coming from the Pauli matrices at each 
vertex in the diagram. For the diagram shown in Fig.~8, 
this factor is $\sum_{i,j} \sigma^i\sigma^j\sigma^i\sigma^j$. 
One can split the sum into the $i=j$ and $i\not=j$ terms, 
and use the fact that $\sigma^i\sigma^i = 1$ and 
$\sigma^i\sigma^j = -\sigma^j\sigma^i$ if $i\not=j$.
Then $\sum_{i,j} \sigma^i\sigma^j\sigma^i\sigma^j = 
\sum_i \sigma^i\sigma^i\sigma^i\sigma^i - \sum_{i\not=j}
\sigma^i\sigma^i\sigma^j\sigma^j = 3 - 6 = -3$. Note 
that the longitudinal spin-fluctuations contribute
a term $\sigma^z\sigma^z\sigma^z\sigma^z = 1$ and thus
the change in sign is caused by the presence of transverse
magnetic modes. The corresponding factor in the case of 
charge-fluctuations is 1, just as in the case without
transverse magnetic modes. Coupling the quasiparticles
to the spin-density instead of the number density produces 
a leading vertex correction with the opposite sign. The 
process depicted in Fig.~8 enhances the magnetic 
interaction\cite{Vertex} while this same diagram 
leads to a suppression of the effective interaction 
in the case of charge-fluctuations. For the range of 
parameters studied, the Monte Carlo simulations 
show that this qualitative difference between coupling 
to magnetic or charge fluctuations persists to all orders, 
save one case, ${\bf k} = (3\pi/8,3\pi/8)$ and $\kappa^2 = 4$, 
where vertex corrections enhance the charge-mediated 
interaction.

As noted above, for large values of $\kappa^2$, 
vertex corrections do not produce qualitative changes, 
but merely quantitative ones. One might therefore ask
whether one can obtain the non-perturbative results 
with a one-loop calculation provided the
parameters of the theory are renormalized.
To illustrate the point, let us focus on the case of 
antiferromagnetic paramagnons with $\kappa^2 = 2$. 
One essentially has two parameters one can renormalize, 
the dimensionless coupling constant $g^2\chi_0/t$ 
and the inverse correlation length $\kappa$. Since 
vertex corrections make the spectral function broader 
and the quasiparticle residue smaller, one could attempt 
to fit the non-perturbative calculations with a one-loop 
theory with a larger coupling constant $g^2\chi_0/t$ 
or smaller $\kappa^2$, or a combination of both. 
If the corrections to the one-loop theory are
purely local, it is possible to absorb 
them in a redefinition of the coupling constant
$g^2\chi_0/t$. It turns out that increasing the
coupling constant does make the quasiparticle residue 
smaller but the one-loop spectral function remains too 
sharp relative to the non-perturbative calculation. By 
the time $g^2\chi_0/t$ is large enough for the 
quasiparticle lifetime to be approximately that 
obtained with the Monte Carlo simulation, the one-loop 
quasiparticle residue is then too small. One has more 
success with making $\kappa^2$ smaller and Fig.~9
compares the spectral function of the one-loop 
calculation with $\kappa^2 = 0.25$ with the 
$A({\bf k},\omega)$ obtained from the Monte Carlo
simulations at $\kappa^2 = 2$. One can get a decent
fit near the hot spot at ${\bf k} = (\pi,\pi/8)$
but the one-loop spectral function is always more
anisotropic in momentum than the non-perturbative 
$A({\bf k},\omega)$ for this value of $\kappa^2$.

The most important feature of the calculations presented
here is the qualitative change in the quasiparticle spectral 
function that occurs as the magnetic interaction gets stronger, 
either on the border of long-range antiferromagnetic or
ferromagnetic order. The appearance of a pseudogap in
the quasiparticle spectrum near a second-order phase 
transition as the temperature approaches the critical 
transition temperature from above was demonstrated for 
the half-filled Hubbard model\cite{Vilk1,Vilk2,Moukouri}, 
for a superconducting 
instability\cite{Vilk2,Vilk3,Allen1,Allen2,Kyung,Rohe,Eckl} 
and for a Peierls-CDW transition\cite{MS}.
By contrast, in the model studied in this paper, 
the quadratic actions of the dynamical molecular
fields do not exhibit a phase transition 
at all (with $\kappa^2 >0$). It may therefore be 
somewhat surprising that a pseudogap is observed 
in the present calculations as the correlation wavevector 
$\kappa$ becomes of order one in the nearly 
antiferromagnetic case and of order one-half for 
nearly ferromagnetic systems.

The physical origin of the pseudogap was explained in 
Refs.~\cite{Vilk1,Vilk2,Moukouri} and later in Ref.~\cite{MS}. 
Quasiparticles only remain coherent for a finite amount 
of time. When the distance they can travel 
during that time becomes shorter than the correlation 
length of the molecular field, quasiparticles effectively 
see long-range order. In the presence of long-range 
antiferromagnetic order, the spin-density-wave quasiparticle 
spectral function consists of two peaks. In the case of 
long-range ferromagnetic order the spin-up quasiparticles 
have an energy shifted downwards, say, relative to the 
paramagnetic quasiparticle energy while the energy of 
a spin-down quasiparticle is shifted upwards relative
to its energy in the absence of long-range ferromagnetic order.
The shifts in energy give rise to corresponding shifts in the
quasiparticle peak in the spectral function. Upon averaging
over the two spin orientations, the quasiparticle spectral 
function would then also consist of two peaks, the up-spin
and down-spin quasiparticle peaks. In our calculations,
when the quasiparticles remain coherent for such a short time
that they effectively see long-range order, the spectral 
functions have charactersitics akin to that of the ordered 
state. For instance, in Fig.~4b, for a coupling of 
quasiparticles to antiferromagnetic spin-fluctuations, 
the spectral function $A({\bf k},\omega)$ for $\kappa^2 = 0.25$ 
looks like that of a broadened spin-density wave. And in the case 
of a ferromagnetically correlated molecular field, in Figs.~7b,c,
$A({\bf k},\omega)$ is "spin-split" as $\kappa^2$ is reduced, 
a feature which can be understood if, during their short 
lifetime, the quasiparticles effectively see ferromagnetic 
order, where the moment is equally likely to point up or down, 
since we are still in the paramagnetic phase.

Quasiparticles only remain coherent for a finite time due
to thermal and quantum fluctuations. The question is what
is that time scale, or the associated characteristic length 
scale to be compared against the magnetic correlation length.
In the half-filled Hubbard model studied in 
Refs.~\cite{Vilk1,Vilk2,Moukouri}, the renormalized classical 
regime for the spin-fluctuations always precedes the zero 
temperature phase transition, and in that regime thermal 
fluctuations dominate and the relevant charactersitic 
length scale is the thermal quasiparticle de Broglie wavelength 
$\xi_{th} = v_F/T$\cite{Vilk1,Vilk2,Moukouri}. Thermal 
fluctuations are dominant near the Peierls-CDW transition 
in two dimensions and the relevant quasiparticle length
scale for the onset of the pseudogap also turns out 
to be $\xi_{th}$\cite{MS}.
For the model considered in this paper, however, thermal
fluctuations do not appear to dominate. Indeed, 
the onset of the pseudogap for coupling of quasiparticles
to antiferromagnetic fluctuations and the qualitative change 
seen in the spectral function for coupling to ferromagnetic 
fluctuations occur for $\xi_{th}\kappa > 1$. The departure
from the criterion $\xi_{th}\kappa\approx 1$ for the breakdown 
of the Midgal approximation is larger in the
ferromagnetic case. The above suggests that another, shorter,
length scale is relevant in the present case. A possible
candidate length scale is the quasiparticle mean-free path. 

One can attempt to make this more quantitative by extracting
quasiparticle lifetimes and mean-free paths from our numerical 
results. In order to do this we fit the quasiparticle peak of
the spectral function with a Lorentzian 
$A_L({\bf k},\omega) = {1\over \pi}
{z_{\bf k}\Gamma_{\bf k}\over (\omega-E_{\bf k})^2 
+ \Gamma^2_{\bf k}}$ which corresponds to a quasiparticle 
approximation for the retarded Green's function 
${\cal G}_R({\bf k},\omega) = {z_{\bf k}\over \omega - E_{\bf k} 
- i\Gamma_{\bf k}}$ describing the propagation of quasiparticles 
of energy $E_{\bf k}$. $\Gamma_{\bf k}$ is related to the
quasiparticle lifetime $\tau_{\bf k}$ through 
$\Gamma_{\bf k}={1\over 2\tau_{\bf k}}$. If one ignores
the momentum dependence of the self-energy, the quasiparticle 
residue $z_{\bf k}$ is related to the ratio of the band
to the effective mass $z_{\bf k} \approx {m\over m^*}$. We 
define the bare velocity $v^{(0)}_{\bf k} = 
\sqrt{({\partial \epsilon_{\bf k}\over \partial k_x})^2+
({\partial \epsilon_{\bf k}\over \partial k_y})^2}$  where
$\epsilon_{\bf k}$ is the band dispersion relation and the
renormalized velocity $v_{\bf k} = z_{\bf k}v^{(0)}_{\bf k}$.
The quasiparticle mean-free path is then approximately
$l_{\bf k} = v_{\bf k}\tau_{\bf k}$. The values of 
$l_{\bf k}$ one obtains for charge-fluctuations as well
as antiferromagnetic and ferromagnetic spin-fluctuations
are shown in Fig.~10. In the case of a coupling of quasiparticles
to charge-fluctuations, Fig.~10a, $l_{\bf k} \gg 1/\kappa$
for all the values of $\kappa$ considered. In this case,
the quasiparticles travel far enough during their lifetime
to see there is no long-range charge order. In the case
of antiferromagnetic spin-fluctuations, Fig.~10b, one sees
that at ${\bf k} = (\pi,\pi/8)$ $l_{\bf k} \sim 1/\kappa$
at $\kappa^2 = 1$, and that is where the pseudogap begins
to appear in the spectral function, Fig.~4b. 
At momentum ${\bf k} = (3\pi/8,3\pi/8)$, the mean-free
path $l_{\bf k} > 1/\kappa$ at $\kappa^2 = 1$ and one 
wouldn't expect a pseudogap. $l_{\bf k}$ becomes less 
than the correlation length at the lowest value of 
$\kappa^2$ and one can see hints of a developing pseudogap
in the spectral function, Fig.~4c. In the ferromagnetic
case, Fig.~10c, the appearance of the two peaks in the
spectral function is broadly consistent with the condition
$l_{\bf k} < 1/\kappa$. Note that we are using a simple
criterion, in an attempt to capture the essential aspects
of the problem, to understand the emergence of a pseudogap 
in our calculations. The crossover to the new state is 
likely to depend on other details not taken into account
by our criterion and is therefore not expected to occur 
exactly at $l_{\bf k} = 1/\kappa$. Moreover, the tails 
in most of the spectral functions are not Lorentzians and 
thus our definition of the quasiparticle lifetime
is clearly approximate. 

We have also extracted mean-free paths for the one-loop 
approximation using the same methodology (results not shown). 
For low enough values of $\kappa^2$, one can get in the 
regime where $l_{\bf k} \leq 1/\kappa$, but no pseudogap
in the spectral function is observed. What is the one-loop
approximation missing? A candidate explanation, based on
an analogy, is the following. When treating the potential 
scattering of a particle in a momentum space basis, to lowest order
of perturbation theory (Born approximation), one assumes the 
wavefunction is unchanged, i.e remains a plane-wave. In 
order to study bound states, in which the wavefunction
of the particle is qualitatively different since it is
localized, one must treat the scattering events to all 
orders. By analogy, the Eliashberg approximation does
not seem to allow for a change in the quasiparticle 
wavefunctions which essentially remain plane-waves. 
In the pseudogap state, the quasiparticle
wavefunctions must be qualitiatively different and
one must allow the quasiparticles to scatter multiple
times against the locally ordered molecular field to
produce the required changes in their wavefunctions.
It seems one would need to sum an infinite set 
of spin-fluctuation exchanges.

The "quenched approximation" formulation of the magnetic 
interaction model actually goes beyond a diagrammatic 
perturbation expansion. The Green's function, Eq.~(\ref{gM}),
depends on the position, spin, and imaginary time, 
but also on the parameters of the theory, 
$\lambda \equiv g^2\chi_0/t$, $\kappa^2$, etc., ${\cal G} = 
{\cal G}(i\sigma\tau ; j\sigma'\tau'|\lambda,\kappa^2,\dots)$.
(To make the dependence on $\lambda = g^2\chi_0/t$ 
explicit, one simply needs to introduce scaled molecular fields 
${\bf m} = g {\bf M}$. The variance of the new variables is 
then $g^2\alpha({\bf q},i\nu_n) \propto g^2\chi({\bf q},i\nu_n)$).
It is clear from Eqs.~(\ref{gM},\ref{ProbM},\ref{NormM}) 
that the theory does not make sense for $\lambda <0$, since 
the Gaussian distributions of the molecular fields would 
have a negative variance. This observation means that 
it is not possible to analytically continue ${\cal G}$ 
to negative values of $\lambda$, implying an essential 
singularity at $\lambda=0$. The perturbation expansion
in powers of $\lambda=0$ is then an asymptotic rather
than convergent series\cite{Negele}. This opens the
possibility for phenomena that lie outside of diagrammatic
perturbation theory. Whether the pseudogap state 
found in the numerical simulations reported on in this
paper is precisely one such phenomenon is not presently
known to the author.

The results presented here do not imply one couldn't get 
a pseudogap state when coupling to charge fluctuations.
On the basis of the arguments presented above, if one were
to increase the strength of the charge correlations or the
coupling of quasiparticles to the molecular field $\Phi$,
one would get in the regime $l_{\bf k} \ll 1/\kappa$ and
the spectral function ought to resemble that of a broadened
charge-density-wave state. Our calculations simply show that
one would have to be closer to the ordered state in the
case of charge fluctuations than for magnetic fluctuations,
under otherwise similar conditions.

\section{Outlook}

We studied a non-perturbative formulation of the 
magnetic interaction model, in which quasiparticles 
are coupled to a Gaussian distributed dynamical 
molecular exchange field. Far from the magnetic
boundary, the type of vertex corrections considered
here do not bring about qualitative changes to the
quasiparticle spectrum. But as one gets closer to
the border of long-range magnetic order, we 
find, for a range of model parameters, that Migdal's 
theorem does not apply and the quasiparticle spectrum is 
qualitatively different from its Eliashberg approximation. 
The physical origin of the phenomenon is that if the 
distance quasiparticles can travel during their lifetime
becomes shorter than the molecular field correlation length, 
these quasiparticles effectively see long-range order. When 
the molecular field correlations are antiferromagnetic, 
the quasiparticle spectral function has the two-peak
structure of a spin-density-wave state, even though there 
is no spontaneous symmetry breaking. We find that the 
associated pseudogap is strongly anisotropic in that
it vanishes along the diagonal of the Brillouin zone
and is large near the zone boundary, in qualitative
agreement with photoemission experiments on underdoped
cuprates\cite{Photoemission}. The anisotropy of the 
pseudogap found in our calculations simply reflects the 
anisotropy of the quasiparticle mean-free path. For 
coupling to ferromagnetic fluctuations, we also find a 
range of parameters were the quasiparticle spectral 
function becomes qualitatively different from its 
one-loop self-consistent approximation. The local 
ferromagnetic order also leads to a splitting of the 
quasiparticle peak into two. These pseudogap effects are 
found to be weaker for nearly ferromagnetic systems than 
for their nearly antiferromagnetic counterparts, under 
otherwise similar conditions.

In the standard theory of quantum critical 
phenomena\cite{HertzMillis}, if $d+z<4$ where $d$ 
is the spatial dimensionality and $z$ the dynamical 
exponent, the mode-mode coupling parameter diverges upon 
renormalization as one approaches the instability
and the critical exponents are different from their 
mean-field values. Therefore, if $d+z<4$, one would 
certainly expect Migdal's theorem to qualitatively break 
down close enough to the quantum critical point. However, 
when $d+z>4$, the critical exponents take their mean-field 
values, as in a one-loop calculation. In that case, one 
could therefore expect, and it is often assumed, 
that the Eliashberg approximation is at least qualitatively 
correct. For the magnetic spectrum considered here, 
Eq.~(\ref{chiML}), $z=2$ for antiferromagnetic fluctuations 
and $z=3$ for ferromagnetic fluctuations. Hence $d+z=4$, 
the marginal dimension in the case of antiferromagnetic 
spin-fluctuations but crucially, $d+z=5$ in the case
of ferromagnetic spin-fluctuations. One could thus
have expected that in the latter case, the one-loop 
self-consistent calculations should be at least 
qualitatively correct since $d+z>4$. This is at variance
with our results, which show qualitative differences
bewteen the non-perturbative calculations and the
Eliashberg predictions. Note that the qualitative
break down of Migdal's theorem occurs for smaller
value of $\kappa^2$ in the ferromagnetic case than
for antiferromagnetic spin-fluctuations. In that sense
the effect of vertex corrections is weaker when
$d+z=5$ compared to the case $d+z=4$, which is what 
is expected.

To summarize, an often assumed criterion for the 
qualitative applicability of the Eliashberg theory near 
an instability, namely $d+z>4$, is clearly a necessary 
condition but the calculations presented here show 
that it is not a sufficient one. This result may point 
to certain limitations of the standard theory of quantum 
critical phenomena\cite{HertzMillis}, which has recently
been criticized by Anderson\cite{PWA}.

The crucial role of dimensionality for pseudogap 
phenomena has been emphasized by Tremblay and 
collaborators\cite{Vilk1,Vilk2,Allen1}
and by Preosti et al.\cite{Preosti}. Since
critical fluctuations responsible for the precursor 
pseudogap are much stronger in two dimensions than 
for three dimensional systems, they find that pseudogap 
phenomena are much weaker in three dimensions. 
One would expect similar results for the magnetic 
interaction model studied in the present paper. 
The role of lattice anisotropy 
is already important at the Eliashberg 
level\cite{Nakamura,Arita,ML2,ML3}, 
because of the increased phase space of soft 
magnetic fluctuations in lower dimensional systems. 
This phase space argument leads one to expect weaker
vertex corrections to the Eliashberg theory of the
magnetic interaction model, and hence weaker pseudogap
effects in isotropic three dimensional systems.

The results presented here raise a number of obvious
questions, which we hope to answer in the future. For 
instance, what are the corrections to the Eliashberg 
theory for the superconducting transition temperature 
to the $d_{x^2-y^2}$ pairing state? We have studied
corrections to quasiparticle spectral properties, 
and found that vertex corrections can produce a 
pseudogap in the quasiparticle spectrum. When the
magnetic correlations are weak enough vertex corrections
do not bring about qualitative changes but nevertheless
reduce the quasiparticle lifetime. The above effects are 
expected to suppress magnetic pairing. 
But one must also include in the calculation of the 
superconducting transition temperature the corresponding 
corrections to the pairing interaction.
In leading order, it was shown\cite{Vertex} that vertex 
corrections lead to a stronger pairing interaction in
the d-wave channel. There may therefore be some 
cancellation of errors, at least in some range of model
parameters. At this stage, however, one can only speculate
about the effect of vertex corrections on the Eliashberg
theory of the superconducting transition temperature.

Transport and thermodynamic properties in the quenched 
approximation would also be of great interest. The
calculations in Ref.\cite{MS} showed that the pseudogap
in the single particle density of states also appeared
in the two-particle Green's functions such as the
optical conductivity and the uniform Pauli susceptibility.
I therefore expect that the pseudogap observed in the
quasiparticle spectral function will show up in the
thermodynamic and transport properties of the model
studied here. The extent to which the model is able
to explain the normal state experimental data on
the underdoped cuprates is an open question.

Our simplifying assumptions, which were born out of 
the necessity to carry out the calculations in a 
reasonable amount of time, should be relaxed. The 
dynamical molecular field correlation function which 
enters their distribution was taken to be of the same
functional form for all the calculations. In Nature, 
however, the distribution function of the molecular 
field experienced by one quasiparticle is determined 
self-consistently by all of the other quasiparticles 
in the system. It is therefore expected to change as 
the nature of the quasiparticle spectrum changes. One 
should also study the effect of mode-mode coupling terms 
which were ignored in the present study. Monien has
recently carried out such a study for a model
of the one-dimensional Peierls-CDW\cite{Monien}.

There is no doubt that the extension of the theory to
deal with effects ignored here will bring about quantitative 
changes to our results. It will be of great interest to 
figure out if and for what model parameters our results are 
modified qualitatively. I would like to think that the
physical origin of the emergence of a pseudogap in the 
quasiparticle spectrum will turn out to be independent 
of the details of the model.

\section{Acknowledgments}

I would like to thank P. Coleman, J.R. Cooper, P.B. Littlewood, 
G.G. Lonzarich, J. Loram, and D. Pines for discussions on this 
and related topics. We acknowledge the support of the EPSRC, 
the Newton Trust and the Royal Society.

\section{Appendix: Green's functions}

In this appendix we derive the mathematical formulas for the 
single-particle Green's functions in a fluctuating exchange or 
scalar field. We also give the algorithm for the numerically stable 
and efficient calculation of such Green's functions. 
We shall make use of fermion coherent states and will derive the 
path-integral for an anti-normal ordered\cite{Patrick} Hamiltonian 
quadratic in the fermion field operators.

We first need an identity for expressing the exponential of
the quadratic form

\begin{equation}
\hat{Q}_A = \sum_{i,j} \psi_i A_{i,j} \psi^\dagger_j
\label{quadA}
\end{equation}

\noindent in terms of fermion coherent states. In the above expression,
$A = A^\dagger$ is a Hermitian matrix of dimension N 
and $\{\psi\},\{\psi^\dagger\}$ are fermion operators. One has:

\begin{equation}
\exp\Big(\sum_{i,j} \psi_i A_{i,j} \psi^\dagger_j\Big) = 
\int [\prod_{i=1}^N d\xi^*_i d\xi_i] |\xi>
\exp\Big(\sum_{i,j}\xi_i (e^A)_{i,j} \xi^*_j\Big)<\xi|
\label{cstates1}
\end{equation}

\noindent where $\{\xi^*\},\{\xi\}$ are Grassmann numbers.
This identity is proved by going into a basis in which the matrix 
$A$ is diagonal. Let $U$ be the unitary transformation that diagonalizes
$A$

\begin{equation}
U^\dagger A U = \Lambda \equiv {\rm diag}(\lambda_1,\dots,\lambda_N)
\label{proof1}
\end{equation}

In terms of the new fermion fields $\{\Phi\},\{\Phi^\dagger\}$

\begin{eqnarray}
\psi_i & = & \sum_{j} U^*_{i,j}\Phi_j\\
\psi^\dagger_i & = & \sum_{j} U_{i,j}\Phi^\dagger_j\\
\Phi_i & = & \sum_{j} U_{j,i}\psi_j \\
\Phi^\dagger_i & = & \sum_{j} U^*_{j,i}\psi^\dagger_j
\end{eqnarray}

\noindent the quadratic form Eq.~(\ref{quadA}) becomes 
$\hat{Q}_A = \sum_i \lambda_i\Phi_i\Phi^\dagger_i$. Since 
$(\Phi_i\Phi^\dagger_i)^n = \Phi_i\Phi^\dagger_i$ for 
$n=1,2,\dots,\infty$, and the terms with different
indices $i$ commute with one another, one has

\begin{equation}
\exp(\hat{Q}_A) = \prod_i \exp(\lambda_i\Phi_i\Phi^\dagger_i)
= \prod_i \Big\{1 + [\exp(\lambda_i)-1]\Phi_i\Phi^\dagger_i\Big\}
\label{equadA}
\end{equation}

\noindent Expanding the product into sums yields

\begin{eqnarray}
\exp(\hat{Q}_A)  & = & 1 + \sum_{i_1} [\exp(\lambda_{i_1})-1]
\Phi_{i_1}\Phi^\dagger_{i_1} + \sum_{i_1 < i_2}
[\exp(\lambda_{i_1})-1]
[\exp(\lambda_{i_2})-1]\Phi_{i_1}\Phi^\dagger_{i_1}
\Phi_{i_2}\Phi^\dagger_{i_2} \nonumber \\
& + & \sum_{i_1 < i_2 < i_3}[\exp(\lambda_{i_1})-1]
[\exp(\lambda_{i_2})-1]
[\exp(\lambda_{i_3})-1]\Phi_{i_1}\Phi^\dagger_{i_1}
\Phi_{i_2}\Phi^\dagger_{i_2}\Phi_{i_3}\Phi^\dagger_{i_3} 
+ \dots \nonumber \\
& + &\sum_{i_1 < \dots < i_N}[\exp(\lambda_{i_1})-1]\dots 
[\exp(\lambda_{i_N})-1] \Phi_{i_1}\Phi^\dagger_{i_1} \dots 
\Phi_{i_N}\Phi^\dagger_{i_N}
\label{equadA2}
\end{eqnarray}

\noindent Since in the various sums one has 
$i_1 \not= i_2 \not= i_3 \dots \not= i_N$, one can
straightforwardly anti-normal order the terms

\begin{eqnarray}
\Phi_{i_1}\Phi^\dagger_{i_1}\Phi_{i_2}\Phi^\dagger_{i_2} & = &
\Phi_{i_2}\Phi_{i_1}\Phi^\dagger_{i_1}\Phi^\dagger_{i_2} \\
\Phi_{i_1}\Phi^\dagger_{i_1}
\Phi_{i_2}\Phi^\dagger_{i_2}\Phi_{i_3}\Phi^\dagger_{i_3} & = &
\Phi_{i_3}\Phi_{i_2}\Phi_{i_1}\Phi^\dagger_{i_1}\Phi^\dagger_{i_2}
\Phi^\dagger_{i_3} \\
\Phi_{i_1}\Phi^\dagger_{i_1} \Phi_{i_2}\Phi^\dagger_{i_2}\dots 
\Phi_{i_N}\Phi^\dagger_{i_N} & = & \Phi_{i_N}\Phi_{i_{N-1}}\dots\Phi_{i_1}
\Phi^\dagger_{i_1}\Phi^\dagger_{i_2}\dots\Phi^\dagger_{i_N}
\end{eqnarray}

\noindent We can now insert a resolution of the identity 
between the annihilation and creation operators. For instance 

\begin{eqnarray}
\Phi_{i_3}\Phi_{i_2}\Phi_{i_1}\Phi^\dagger_{i_1}\Phi^\dagger_{i_2}
\Phi^\dagger_{i_3} & = & \Phi_{i_3}\Phi_{i_2}\Phi_{i_1} 
\int [\prod_{i=1}^N d\xi^*_i d\xi_i] |\xi>
\exp\Big(-\sum_{i}\xi^*_i\xi_i\Big)<\xi|
\Phi^\dagger_{i_1}\Phi^\dagger_{i_2}\Phi^\dagger_{i_3} \nonumber \\
& = & \int \prod_{i=1}^N d\xi^*_i d\xi_i \xi_{i_3}\xi_{i_2}
\xi_{i_1}|\xi>\exp\Big(-\sum_{i}\xi^*_i\xi_i\Big)
<\xi|\xi^*_{i_1}\xi^*_{i_2}\xi^*_{i_3} \nonumber \\
& = & \int [\prod_{i=1}^N d\xi^*_i d\xi_i]\; \xi_{i_1}\xi^*_{i_1}
\xi_{i_2}\xi^*_{i_2}\xi_{i_3}\xi^*_{i_3}
\exp\Big(-\sum_{i}\xi^*_i\xi_i\Big)|\xi><\xi| \nonumber
\end{eqnarray}

\noindent Applying the above operation to every term in the sum in 
Eq.~(\ref{equadA2}) one obtains

\begin{eqnarray}
\exp(\hat{Q}_A) & = &\int [\prod_{i=1}^N d\xi^*_i d\xi_i] |\xi>
\exp\Big(-\sum_{i}\xi^*_i\xi_i\Big)
\prod_{i=1}^N [1 + (e^{\lambda_i}-1)\xi_i\xi^*_i]<\xi| \nonumber \\
& = & \int [\prod_{i=1}^N d\xi^*_i d\xi_i] |\xi>
\exp\Big(-\sum_{i}\xi^*_i\xi_i\Big)
\exp\Big(\sum_{i}(e^{\lambda_i}-1)\xi_i\xi^*_i\Big)<\xi|\nonumber \\
& = & \int [\prod_{i=1}^N d\xi^*_i d\xi_i] |\xi>
\exp\Big(\sum_{i}e^{\lambda_i}\xi_i\xi^*_i\Big)<\xi|
\label{equadA3}
\end{eqnarray}

\noindent Going back to the original representation in which 
the matrix A is not diagonal yields Eq.~(\ref{cstates1}). If 
we have another quadratic form

\begin{equation}
\hat{Q}_B = \sum_{ij} \psi_i B_{i,j} \psi^\dagger_j
\label{quadB}
\end{equation}

\noindent where $B = B^\dagger$ is a Hermitian matrix of 
dimension N, it is a simple exercise in Grassmann integration 
to show that

\begin{equation}
\exp(\hat{Q}_A)\exp(\hat{Q}_B) = \int [\prod_{i=1}^N d\xi^*_i d\xi_i] |\xi>
\exp\Big(\sum_{ij}\xi_i (e^Ae^B)_{i,j} \xi^*_j\Big)<\xi|
\label{cstates2}
\end{equation}

\noindent To prove the above identity one uses the coherent state 
representation Eq.~(\ref{cstates1}) for $\exp(\hat{Q}_A)$ and 
$\exp(\hat{Q}_B)$

\begin{eqnarray}
\exp(\hat{Q}_A) & = & \int [\prod_{i=1}^N d\xi^*_i d\xi_i] |\xi>
\exp\Big(\sum_{ij}\xi_i (e^A)_{i,j} \xi^*_j\Big)<\xi| \\
\exp(\hat{Q}_B) & = & \int [\prod_{i=1}^N d\eta^*_i d\eta_i] |\eta>
\exp\Big(\sum_{ij}\eta_i (e^B)_{i,j} \eta^*_j\Big)<\eta|
\end{eqnarray}

\noindent Integrating over the Grassmann variables $\{\xi^*_i\}$ 
and $\{\eta_i\}$ yields Eq.~(\ref{cstates2}).

The imaginary time single-particle Green's function in a 
time dependent field is given by the usual expression

\begin{eqnarray}
G(i\sigma\tau ; j\sigma'\tau') & = & -\big<T_\tau\{\psi_{i\sigma}(\tau)\
\psi^\dagger_{j\sigma'}(\tau')\}\big> \nonumber \\
& = & -{1\over {\cal Z}}
Tr\Big[T_\tau\Big\{\exp\Big(-\int_0^\beta{\hat h}(s)ds\Big)
\psi_{i\sigma}(\tau)\psi^\dagger_{j\sigma'}(\tau')\Big\}\Big]
\label{green2} \\
{\cal Z} & = & Tr\Big[T_\tau\exp\Big(-\int_0^\beta{\hat h}(s)ds\Big)\Big]
\label{Z}
\end{eqnarray}

\noindent where the time dependent Hamiltonian ${\hat h}(\tau)$ 
is given by Eq.~(\ref{ham1}) and Eq.~(\ref{ham2}) for coupling 
to a fluctuating exchange or scalar field respectively. 
The trace is over the fermion fields. For ease of notation, 
we have dropped the explicit dependence of 
$G(i\sigma\tau ; j\sigma'\tau')$ on the molecular 
field. Because of the time ordering operator, one must distinguish 
the two cases $\tau > \tau'$ and $\tau \leq \tau'$.

\begin{eqnarray}
G(i\sigma\tau ; j\sigma'\tau') & = & -{1\over {\cal Z}} Tr\Bigg[
\Big\{T_\tau\exp\Big(-\int_{\tau}^{\beta}{\hat h}(s)ds\Big)\Big\}
\psi_{i\sigma}(\tau)
\Big\{T_\tau\exp\Big(-\int_{\tau'}^{\tau}{\hat h}(s)ds\Big)\Big\}
\psi^\dagger_{j\sigma'}(\tau') \nonumber \\
& \times &\Big\{T_\tau\exp\Big(-\int_0^{\tau}{\hat h}(s)ds\Big)\Big\} \Bigg]
\quad\quad \tau > \tau'
\label{green3} \\
G(i\sigma\tau ; j\sigma'\tau') & = & -{1\over {\cal Z}} Tr\Bigg[
\Big\{T_\tau\exp\Big(-\int_{\tau'}^{\beta}{\hat h}(s)ds\Big)\Big\}
\psi^\dagger_{j\sigma'}(\tau')
\Big\{T_\tau\exp\Big(-\int_{\tau}^{\tau'}{\hat h}(s)ds\Big)\Big\}
\psi_{i\sigma}(\tau) \nonumber \\
& \times &\Big\{T_\tau\exp\Big(-\int_0^{\tau'}{\hat h}(s)ds\Big)\Big\} \Bigg]
\quad\quad \tau \leq \tau'
\label{green4}
\end{eqnarray}

\noindent The imaginary times $\tau$ and $\tau'$ are 
discretized such that $\tau  = \Delta\tau m \quad m=0,1,\dots L_t$, 
$\tau' = \Delta\tau m' \quad m'=0,1,\dots L_t$,  where 
$\Delta\tau = {\beta\over L_t}$. One then uses the Trotter 
break-up for the Hamiltonian in an exchange, Eq.~(\ref{ham1}),
or scalar field, Eq.~(\ref{ham2}), and hence 
$\exp(-\Delta\tau{\hat h}(m\Delta\tau)) \approx 
\exp(-\Delta\tau{\hat h}_1(m\Delta\tau)) 
\exp(-\Delta\tau{\hat h}_0(m\Delta\tau))$. Up to an 
unimportant constant, the anti-normal ordered hopping 
Hamiltonian ${\hat h}_0$, Eq.~(\ref{ham0}) is written as

\begin{eqnarray}
{\hat h}_0 & = & \sum_{i,\sigma ; j\sigma'}\psi_{i\sigma}T_{i\sigma,j\sigma'}
\psi^\dagger_{j\sigma'} 
\label{h0} \\
T_{i\sigma,j\sigma'} & = & (t_{ij} + \mu\delta_{i,j})\delta_{\sigma,\sigma'}
\label{HoppingMat}
\end{eqnarray}

\noindent Similarly, up to constant terms that drop out of the expression
for the single particle Green's function, the Hamiltonian for the coupling 
to the fluctuating field is written as

\begin{eqnarray}
{\hat h}_1(m\Delta\tau) & = & \sum_{i\sigma j\sigma'}\psi_{i\sigma}
A_{i\sigma,j\sigma'}(m)\psi^\dagger_{j\sigma'} 
\label{h1} \\
A_{i\sigma,j\sigma'}(m) & = & {g\over \sqrt{3}}\delta_{i,j}\Big[
 M^z_i(m\Delta\tau)\delta_{\sigma,\uparrow}\delta_{\sigma',\uparrow}
-M^z_i(m\Delta\tau)\delta_{\sigma,\downarrow}
\delta_{\sigma',\downarrow} \nonumber \\
& + & (M^x_i(m\Delta\tau)+iM^y_i(m\Delta\tau))
\delta_{\sigma,\uparrow}\delta_{\sigma',\downarrow}
+ (M^x_i(m\Delta\tau)-iM^y_i(m\Delta\tau))
\delta_{\sigma,\downarrow}\delta_{\sigma',\uparrow}
\Big]
\label{exchange} \\
A_{i\sigma,j\sigma'}(m) & = & g\delta_{i,j}
\delta_{\sigma,\sigma'}\Phi_i(m\Delta\tau)
\label{scalar}
\end{eqnarray}

\noindent where Eq.~(\ref{exchange}) and Eq.~(\ref{scalar}) are for a
fluctuating exchange and scalar field respectively. Making use of 
Eqs.~(\ref{cstates1},\ref{cstates2}), one has

\begin{eqnarray}
\exp(-\Delta\tau{\hat h}(m\Delta\tau)) & \approx &
\exp(-\Delta\tau{\hat h}_1(m\Delta\tau)) 
\exp(-\Delta\tau{\hat h}_0(m\Delta\tau)) \nonumber \\
& = & \int d\mu(\xi) |\xi>
\exp\Big(\sum_{i\sigma j\sigma'}\xi_{i\sigma} [B_{m}]_{i\sigma,j\sigma'} 
\xi^*_{j\sigma'}\Big)<\xi|
\label{exph} \\
B_{m} & = & \exp(-\Delta\tau A(m))\exp(-\Delta\tau T)
\label{bmat}
\end{eqnarray}

\noindent where from now on we use $d\mu(\xi)$ to denote the Grassmann
integration measure

\begin{equation}
d\mu(\xi) \equiv [\prod_{i\sigma} d\xi^*_{i\sigma} d\xi_{i\sigma}]
\label{measure}
\end{equation}

\noindent In the case of coupling to a scalar field

\begin{equation}
[\exp(-\Delta\tau A(m))]_{i\sigma,j\sigma'} = 
\delta_{i,j}\delta_{\sigma\,\sigma'}\exp(-g\Delta\tau\Phi_i(m\Delta\tau))
\label{AmatScal}
\end{equation}

\noindent The case of the exchange field is slightly more complicated:

\begin{eqnarray}
[\exp(-\Delta\tau A(m))]_{i\sigma,j\sigma'} & = &
\delta_{i,j}[\exp(-\Delta\tau a_i(m))]_{\sigma,\sigma'}
\label{AmatExch1} \\
\exp(-\Delta\tau a_i(m)) & = & \pmatrix{c - s{\cal M}^z_i(m\Delta\tau) &
-s{\cal M}^+_i(m\Delta\tau) \cr -s{\cal M}^-_i(m\Delta\tau) & 
c + s{\cal M}^z_i(m\Delta\tau)}
\label{AmatExch2} \\
c & = & \cosh(g\Delta\tau |{\bf M}_i(m\Delta\tau)|/\sqrt{3})
\label{AmatExch3} \\
s & = & \sinh(g\Delta\tau |{\bf M}_i(m\Delta\tau)|/\sqrt{3})
\label{AmatExch4} \\
{\cal M}^z_i(m\Delta\tau) & = &
{M^z_i(m\Delta\tau)\over |{\bf M}_i(m\Delta\tau)|}
\label{AmatExch5} \\
{\cal M}^{\pm}_i(m\Delta\tau) & = &
{M^x_i(m\Delta\tau) \pm iM^y_i(m\Delta\tau)\over |{\bf M}_i(m\Delta\tau)|}
\label{AmatExch6} \\
|{\bf M}_i(m\Delta\tau)| & = & \sqrt{[M^x_i(m\Delta\tau)]^2
+[M^y_i(m\Delta\tau)]^2+[M^z_i(m\Delta\tau)]^2}
\label{AmatExch7} 
\end{eqnarray}

\noindent Matrix multiplication by $\exp(-\Delta\tau A(m))$ is most 
easily done in real space. We have implemented the matrix multiplication
by $\exp(-\Delta\tau T)$ by means of the Fast Fourier transform instead of
the usual checkerboard decomposition. In the discretized imaginary time
version, 

\begin{eqnarray}
T_\tau\exp\Big(-\int_0^\beta{\hat h}(s)ds\Big) &\approx &
\exp(-\Delta\tau {\hat h}(\Delta\tau L_t))\dots
\exp(-\Delta\tau {\hat h}(\Delta\tau)) \nonumber \\
& = &\int d\mu(\xi) |\xi>
\exp\Big(\sum_{i\sigma j\sigma'}\xi_{i\sigma} 
[B_{L_t}B_{L_t-1}\dots B_{1}]_{i\sigma,j\sigma'} 
\xi^*_{j\sigma'}\Big)<\xi|
\label{norm1}
\end{eqnarray}

Note that we choose to approximate the imaginary time integral of
a function in an interval of length $\Delta\tau$ by $\Delta\tau$ 
times the value of the function at the upper limit of the interval.
The calculation of the normalization factor $\cal Z$, Eq.~(\ref{Z}), 
then becomes a simple exercise in Grassmann integration

\begin{eqnarray}
{\cal Z} & = &\int d\mu(\eta)d\mu(\xi)
<-\eta|\xi>\exp\Big(-\sum_{i\sigma}\eta^*_{i\sigma}\eta_{i\sigma} 
+\sum_{i\sigma j\sigma'}\xi_{i\sigma} 
[B_{L_t}B_{L_t-1}\dots B_{1}]_{i\sigma,j\sigma'} 
\xi^*_{j\sigma'}\Big)<\xi|\eta> \nonumber \\
& = & Det[1 + B^T_{1}B^T_{2}\dots B^T_{L_t}] 
= Det[1 + B^T_{0}B^T_{1}\dots B^T_{L_t-1}]
\label{norm2}
\end{eqnarray}

\noindent where $B^T_{m}$ is the transpose of the matrix 
$B_{m}$ and in the last line we have used the periodicity 
of the exchange and scalar fields which implies 
$B_{L_t} = B_{0}$ and multiplied the expression
by $1 = Det[B^T_{0}{B^T_{0}}^{-1}]$. 

The remainder of the calculation of the
Green's function proceeds along similar lines. Consider 
the case $m > m'$. One has

\begin{eqnarray}
T_\tau\exp\Big(-\int_{\tau}^\beta{\hat h}(s)ds\Big) & \approx &
\exp(-\Delta\tau {\hat h}(\Delta\tau L_t)) \dots
\exp(-\Delta\tau {\hat h}(\Delta\tau (m+1))) \nonumber \\
& = & \int d\mu(\xi) |\xi>
\exp\Big(\sum_{i\sigma j\sigma'}\xi_{i\sigma} 
[B_{L_t}B_{L_t-1}\dots B_{m+1}]_{i\sigma,j\sigma'} 
\xi^*_{j\sigma'}\Big)<\xi|
\label{green5}
\end{eqnarray}  

\noindent with similar expressions  for 
$T_\tau\exp\Big(-\int_{\tau'}^{\tau}{\hat h}(s)ds\Big)$ and
$T_\tau\exp\Big(-\int_0^{\tau'}{\hat h}(s)ds\Big)$. After 
straightforward algebraic manipulations, one arrives at:

\begin{eqnarray}
G(i\sigma m,j\sigma' m') & = & -{1\over {\cal Z}}\int 
d\mu(\xi)d\mu(\eta)d\mu(\rho)d\mu(\phi)
\exp\Big(-\sum_{i\sigma}\xi^*_{i\sigma}\xi_{i\sigma}\Big) 
\exp\Big(-\sum_{i\sigma}\xi^*_{i\sigma}\eta_{i\sigma}\Big) 
\nonumber \\
& &\exp\Big(\sum_{i\sigma}\eta^*_{i\sigma}\rho_{i\sigma}\Big)
\exp\Big(\sum_{i\sigma}\rho^*_{i\sigma}\phi_{i\sigma}\Big) 
\rho_{i\sigma}\rho^*_{j\sigma'}
\exp\Big(-\sum_{i\sigma,j,\sigma'} 
\eta_{i\sigma} M^\eta_{i\sigma,j\sigma'}\eta^*_{j\sigma'}\Big)
\nonumber \\
& &\exp\Big(-\sum_{i\sigma j\sigma'} 
\rho_{i\sigma} M^\rho_{i\sigma,j\sigma'}\rho^*_{j\sigma'}\Big)
\exp\Big(-\sum_{i\sigma j\sigma'} 
\phi_{i\sigma} M^\phi_{i\sigma,j\sigma'}\phi^*_{j\sigma'}\Big)
\label{green6} \\
M^\eta & = & B_{L_t}B_{L_t-1}\dots B_{m+1} 
\label{green7} \\
M^\rho & = & B_{m}B_{m-1}\dots B_{m'+1} 
\label{green8} \\
M^\phi & = & B_{m'}B_{m'-1}\dots B_{1} 
\label{green9}
\end{eqnarray}

\noindent The integral over the quadratic form is easily computed. 
The calculation in the case $m \leq m'$ proceeds along similar 
lines and we just quote the final answer. The single particle 
Green's function is given by the following matrix expressions

\begin{equation}
G(m,m') = \cases{-B^T_{m+1}\dots B^T_{L_t}B^T_{1}\dots B^T_{m'}
\Big[1 + B^T_{m'+1}\dots B^T_{L_t}B^T_{1}\dots B^T_{m'}\Big]^{-1} 
&if $m > m'$ \cr
B^T_{m+1}B^T_{m+2}\dots B^T_{m'}\Big[
1 + B^T_{m'+1}\dots B^T_{L_t}B^T_{1}\dots B^T_{m'}\Big]^{-1} 
&if $m \leq m'$\cr}
\label{greenFinal}
\end{equation}

Note that these expressions differ from those of Ref.\cite{BSS} 
since these authors are using the standard formulation of the path 
integral with normal ordered operators, instead of the anti-normal 
ordering used here. The numerical calculation of the single particle 
Green's function, Eq.~(\ref{greenFinal}), is complicated by the fact 
that the matrix products entering the above expressions are 
seriously ill-conditioned\cite{StableG}. I have found that the 
Gram-Schmidt matrix factorization algorithm of Ref.\cite{StableG}
can be unstable at sufficiently low temperatures and for strongly 
fluctuating molecular fields. I therefore use an alternative way 
to compute the Green's function which I have found to be 
particularly stable. It is based on the matrix identities:

\begin{eqnarray}
(1+A_1A_2)^{-1} & = & T_2\Big[1 - T_1 - T_2 + 2T_1T_2\Big]^{-1}T_1
\label{MatIden1} \\
A_2(1+A_1A_2)^{-1} & = & \Big[1 - T_2\Big]
\Big[1 - T_1 - T_2 + 2T_1T_2\Big]^{-1}T_1
\label{matIden2} \\
T_i & = & (1+A_i)^{-1}
\label{Tdef}
\end{eqnarray}

In practice we calculate $G(m,0)$ and $G(0,m')$. The above 
matrix identities can be used recursively to calculate these 
Green's functions in a stable manner. The matrices $A_1,A_2$ 
are taken as products of $B^T$ matrices. The number of $B^T$ 
matrices in a "block" is chosen to be the maximum number such
that $(1+A_1)$, $(1+A_2)$ can be inverted without significant 
loss of precision. One can then obtain the Green's functions 
$G(m,0)$ and $G(0,m')$ at a subset of values $m,m'$. The Green's 
function for the other values of $m,m'$ can be obtained by 
propagating forward or backward in time. For example, 
$G(m+1,0) = [B^T_{m+1}]^{-1}G(m,0)$. We leave it to the 
reader to fill in the details.

\begin{figure}
\centerline{\epsfysize=6.00in
\epsfbox{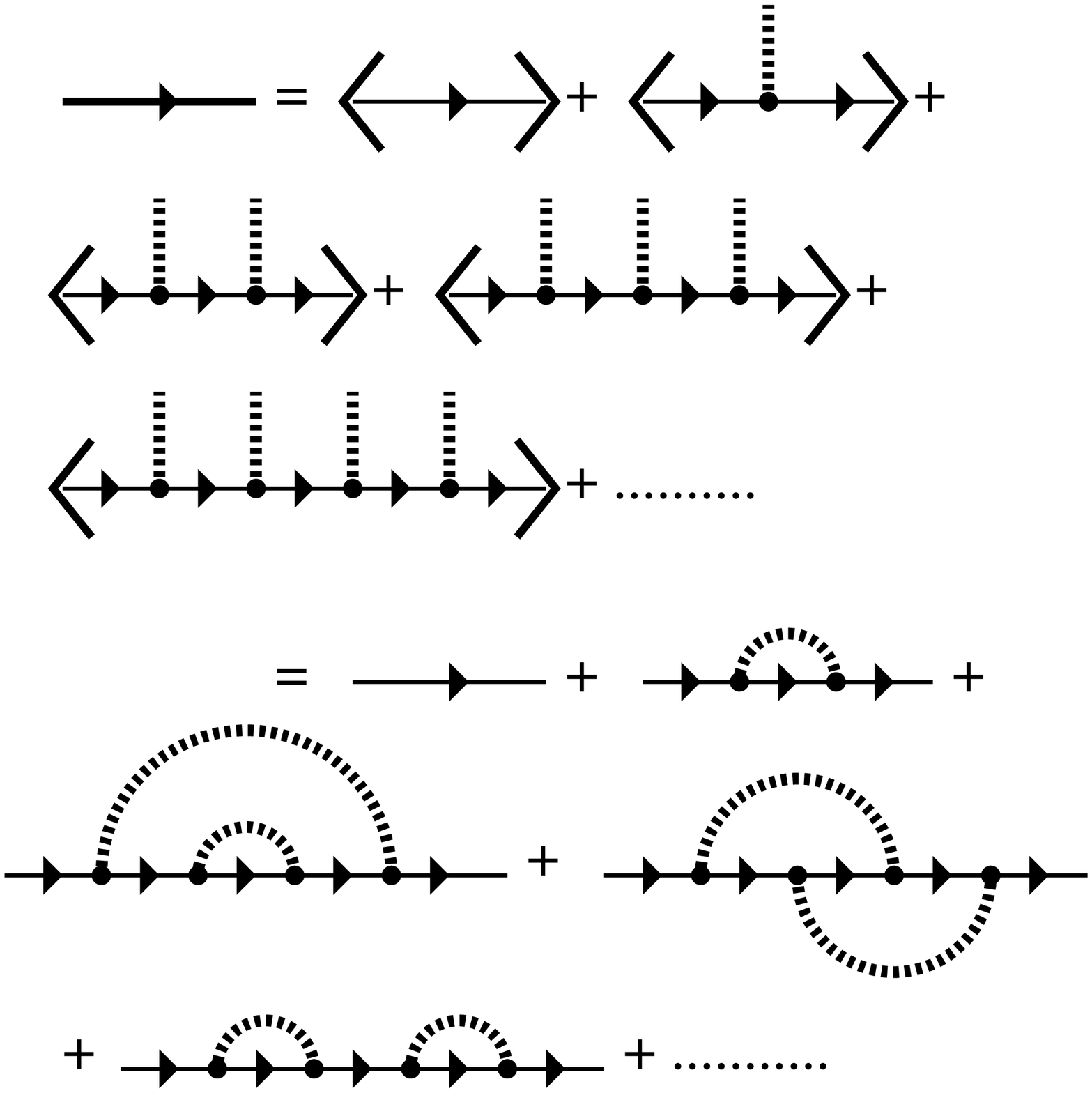}}
\caption{Diagrammatic expansion for the single particle Green's
function. The dashed line connected at one end only to the
solid line represents the interaction of the fermions
with the dynamical molecular field and the brackets the
average over the Gaussian distribution of these fields. 
The averaging over the distribution of molecular fields 
is carried out by pairing the dashed lines in all possible
ways, each pairing giving a factor proportional to the
two-point correlation function of the molecular field,
the dynamical susceptibility $\chi({\bf q},i\nu_n)$
according to Eqs.~(\ref{Mcorr},\ref{Pcorr}). The lower 
part of the figure shows the pairings one obtains up 
to two spin or charge-fluctuation exchanges.}
\label{fig1}
\end{figure}

\begin{figure}
\centerline{\epsfysize=6.00in
\epsfbox{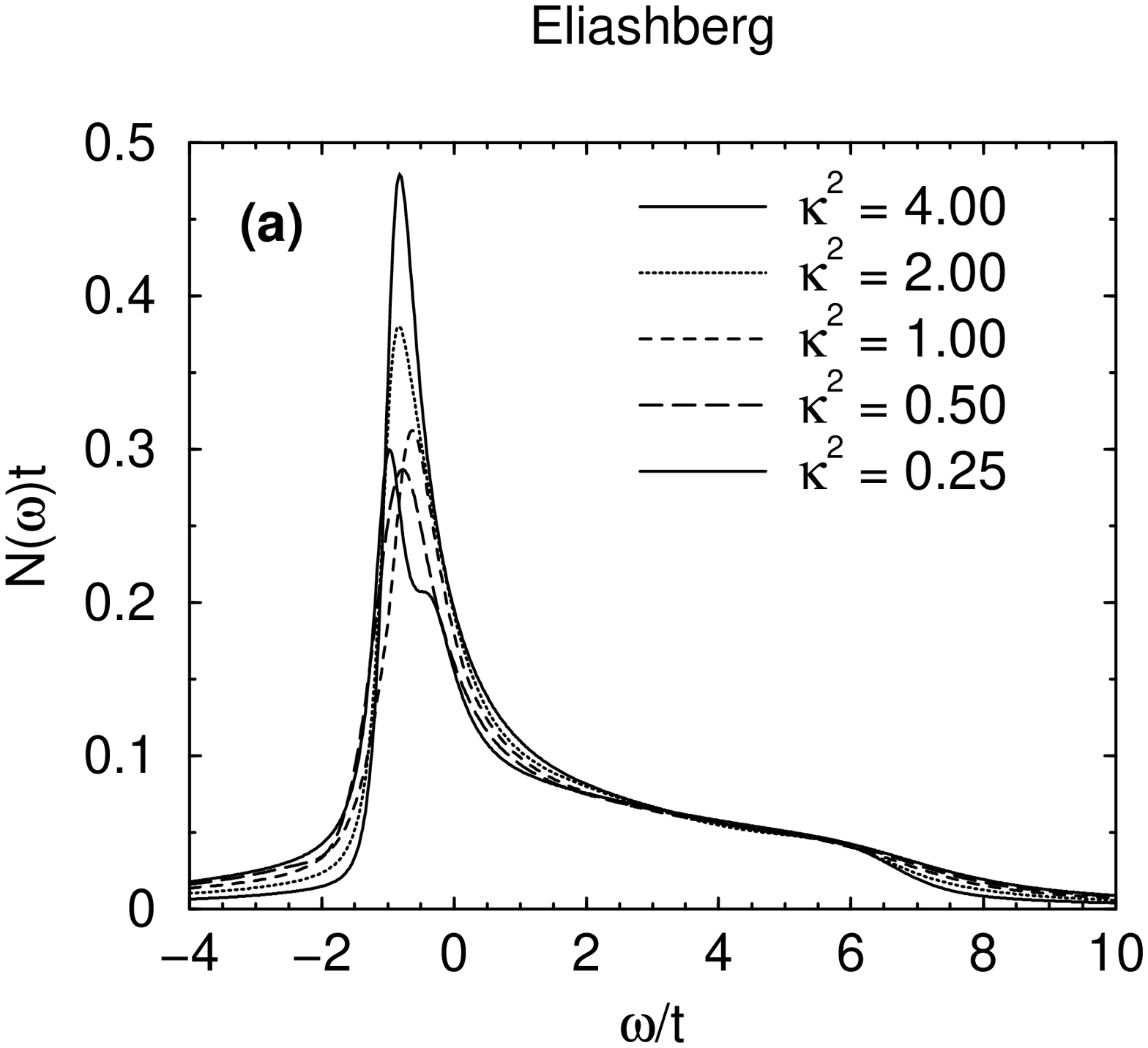}}
\centerline{\epsfysize=6.00in
\epsfbox{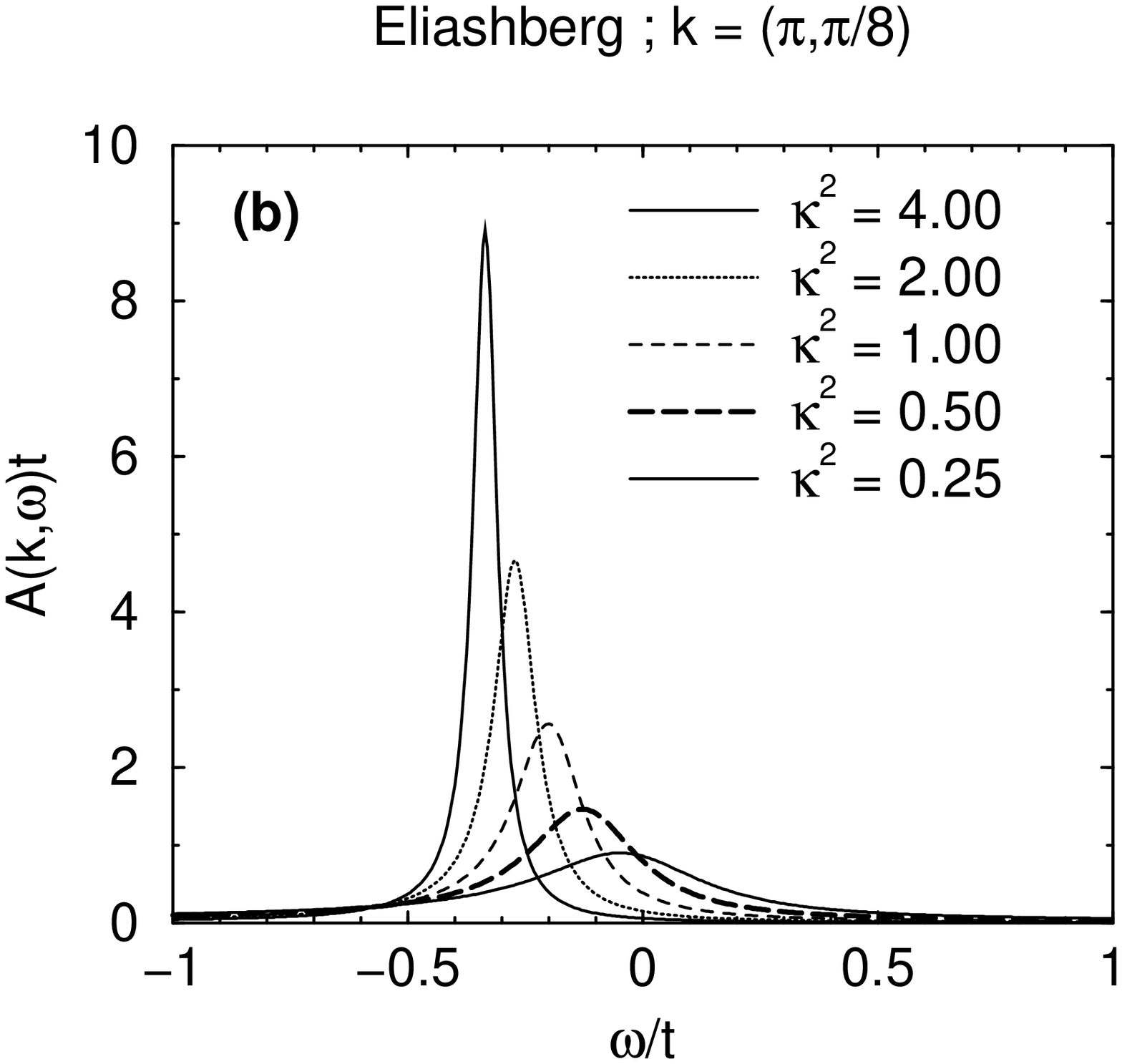}}
\centerline{\epsfysize=6.00in
\epsfbox{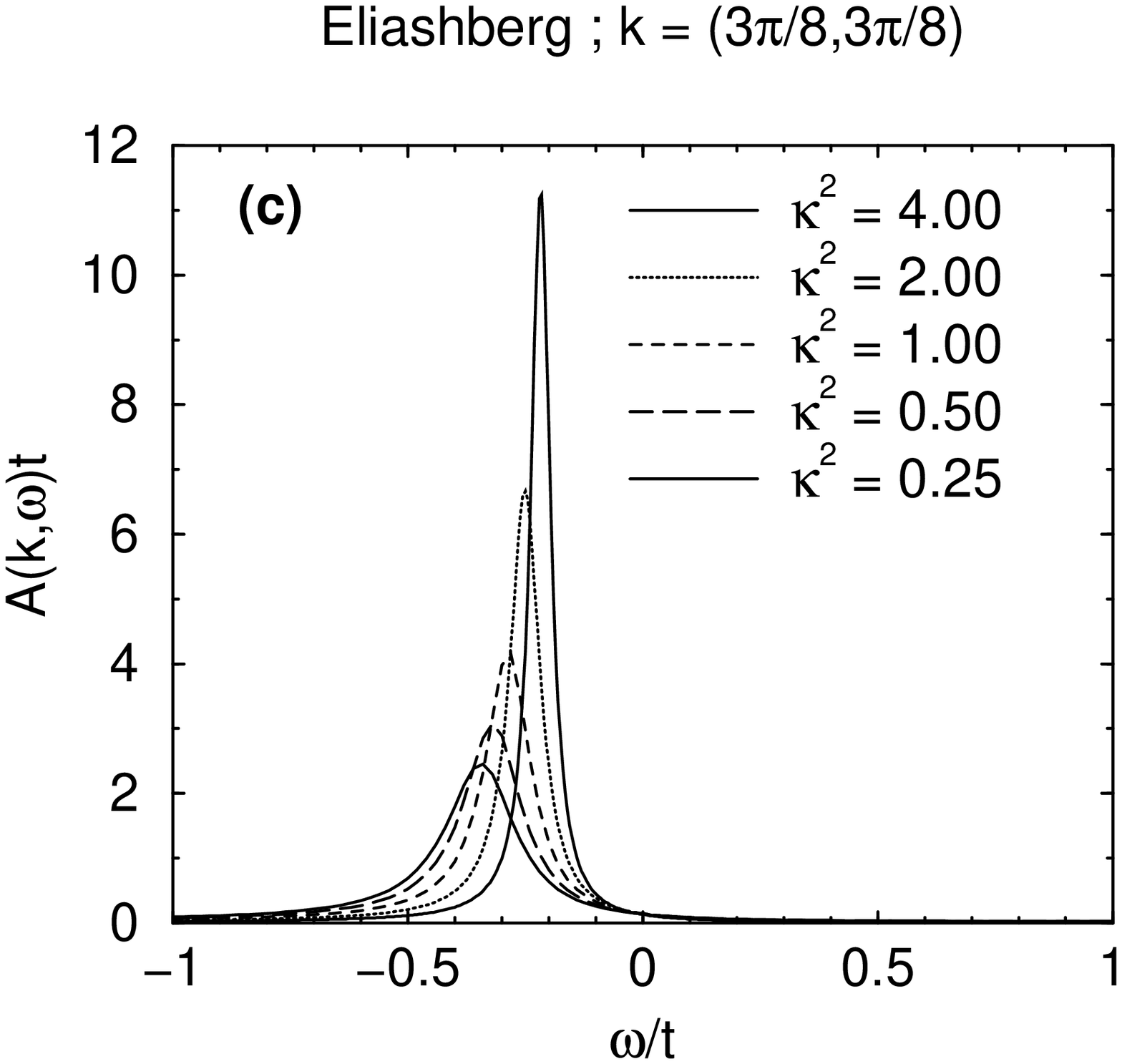}}
\label{fig2}
\caption{The one-loop approximation for the quasiparticle
properties for both commensurate charge-fluctuations and 
antiferromagnetic spin-fluctuations.
The tunneling density of states $N(\omega)$ is shown in (a)
while (b) and (c) show the quasiparticle spectral function 
$A({\bf k},\omega)$ for momenta just below the Fermi level. 
(b) shows $A({\bf k},\omega)$ for a wavevector close to 
the Van Hove singularity and (c) shows $A({\bf k},\omega)$ 
for a wavevector along the diagonal of the Brillouin zone.}
\end{figure}

\begin{figure}
\centerline{\epsfysize=6.00in
\epsfbox{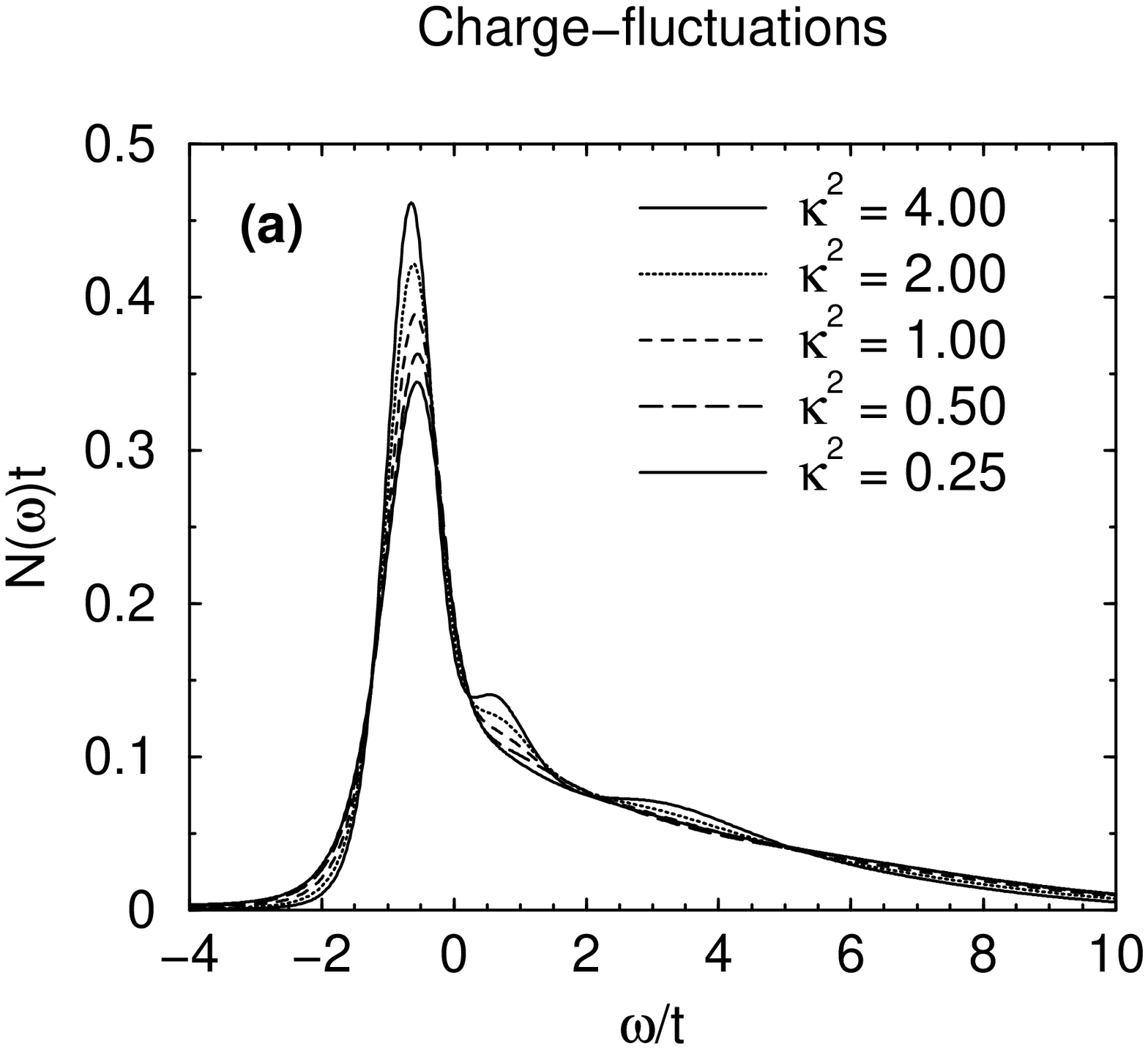}}
\centerline{\epsfysize=6.00in
\epsfbox{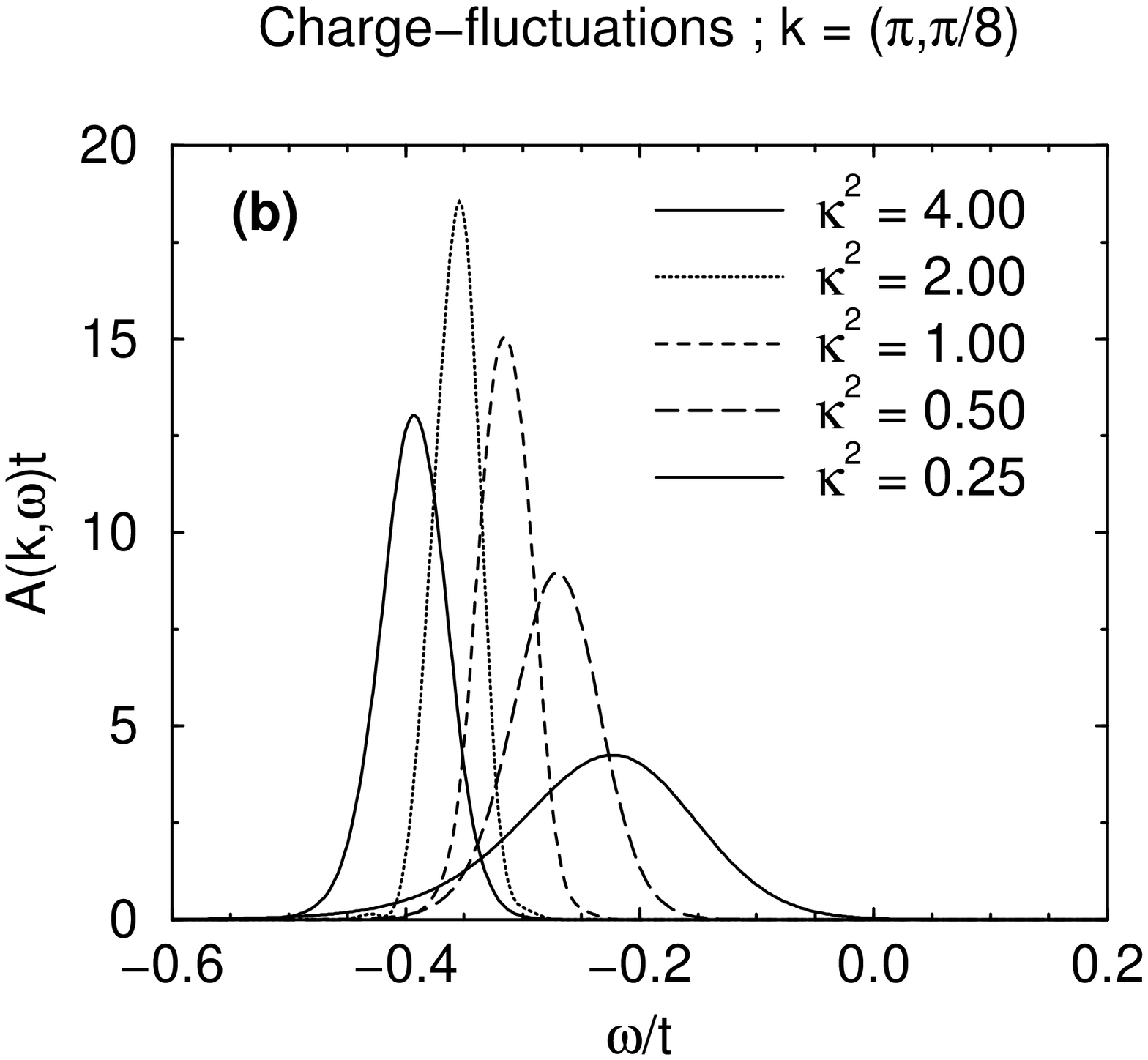}}
\centerline{\epsfysize=6.00in
\epsfbox{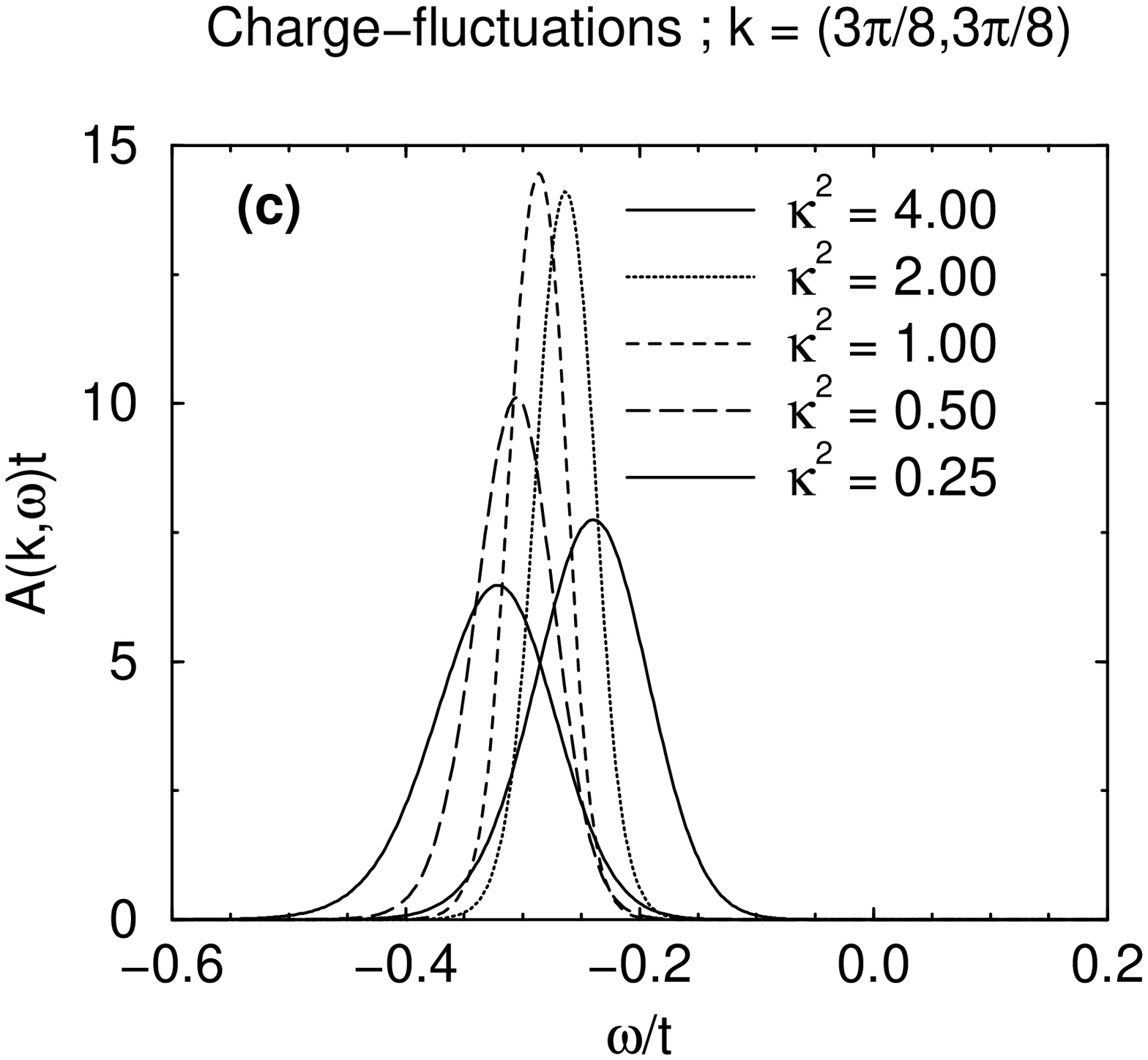}}
\label{fig3}
\caption{Quasiparticle properties for a coupling to a 
scalar dynamical molecular field $\Phi$. The tunneling 
density of states $N(\omega)$ is shown in (a) while (b) 
and (c) show the quasiparticle spectral function 
$A({\bf k},\omega)$ for momenta just below the Fermi level. 
(b) shows $A({\bf k},\omega)$ for a wavevector close to 
the Van Hove singularity and (c) shows $A({\bf k},\omega)$ 
for a wavevector along the diagonal of the Brillouin zone.}
\end{figure}

\begin{figure}
\centerline{\epsfysize=6.00in
\epsfbox{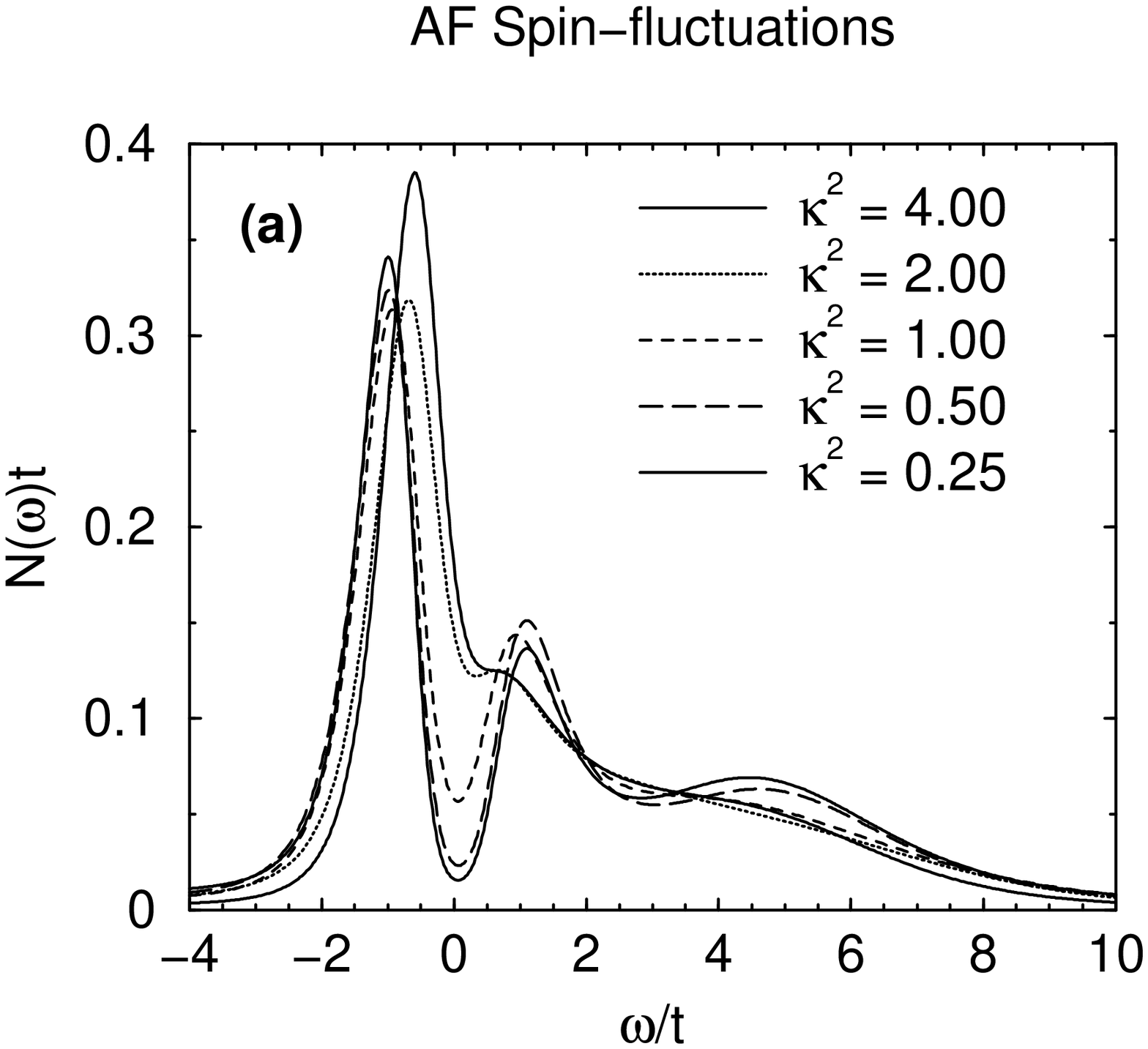}}
\centerline{\epsfysize=6.00in
\epsfbox{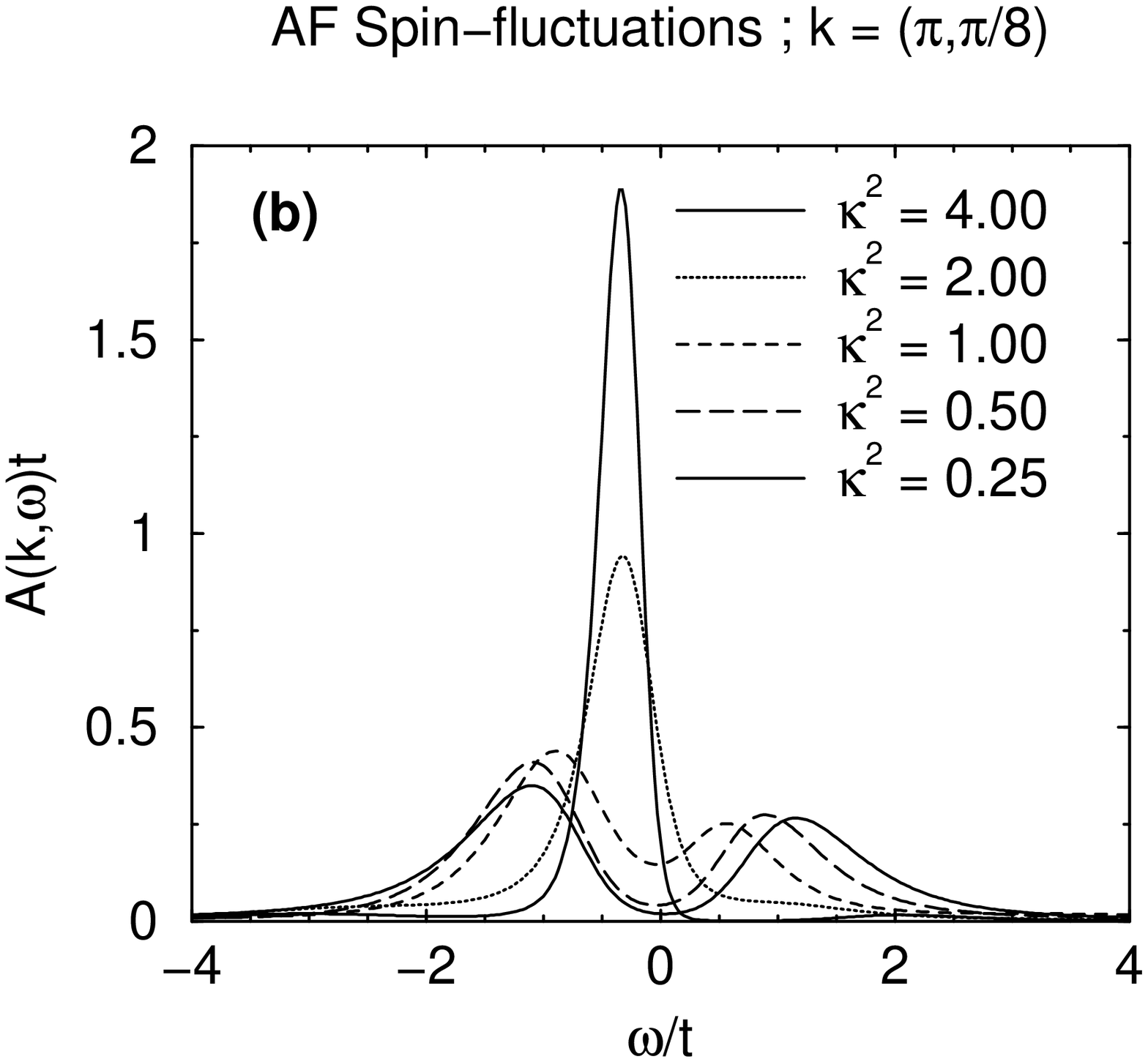}}
\centerline{\epsfysize=6.00in
\epsfbox{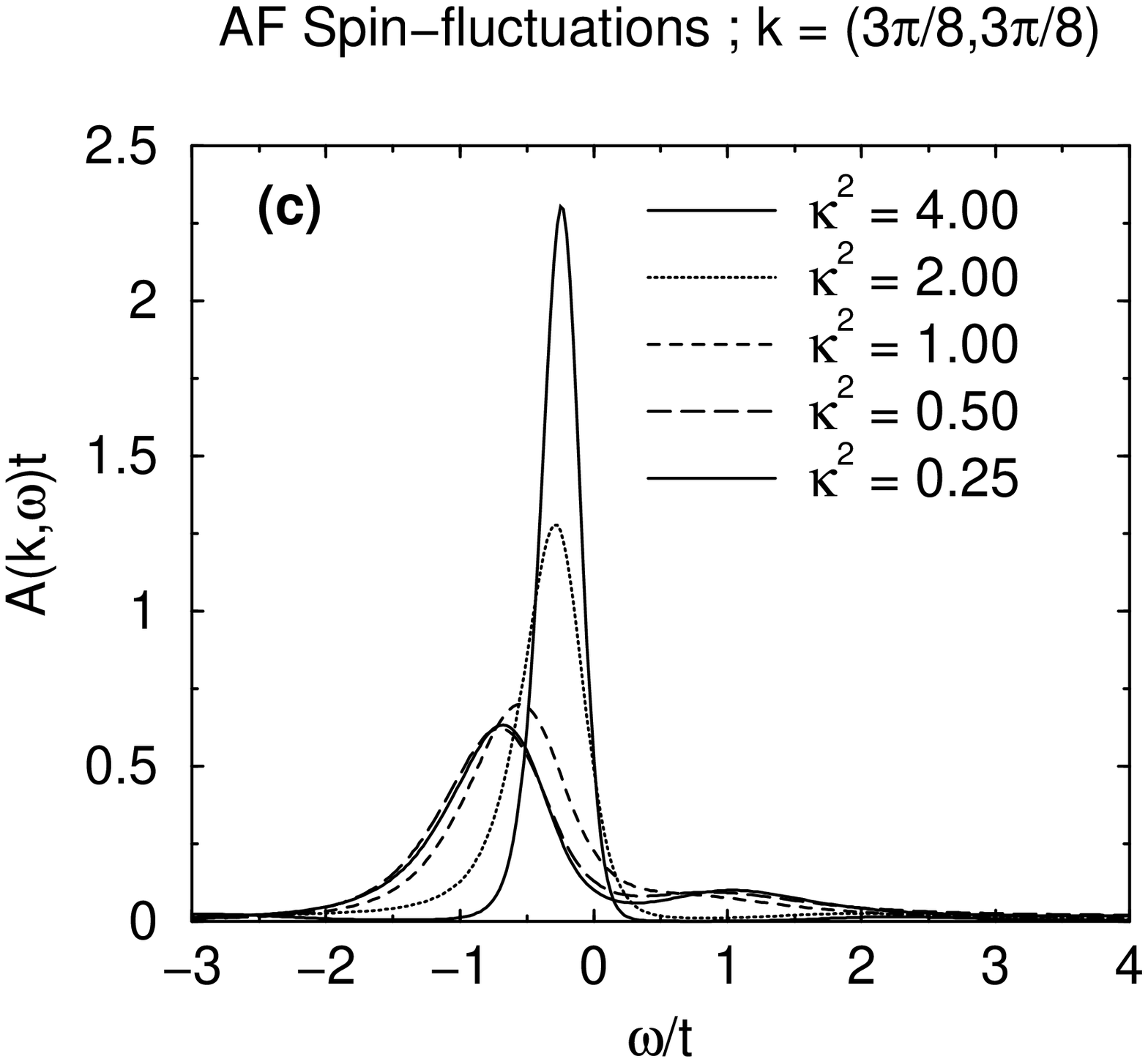}}
\label{fig4}
\caption{Quasiparticle properties for a coupling to an 
exchange vector dynamical molecular field ${\bf M}$
with antiferromagnetic correlations. 
The tunneling density of states $N(\omega)$ is shown in 
(a) while (b) and (c) show the quasiparticle spectral 
function $A({\bf k},\omega)$ for momenta just below the 
Fermi level. (b) shows $A({\bf k},\omega)$ for a wavevector 
close to the Van Hove singularity and (c) shows $A({\bf k},\omega)$ 
for a wavevector along the diagonal of the Brillouin zone.}
\end{figure}

\begin{figure}
\centerline{\epsfysize=6.00in
\epsfbox{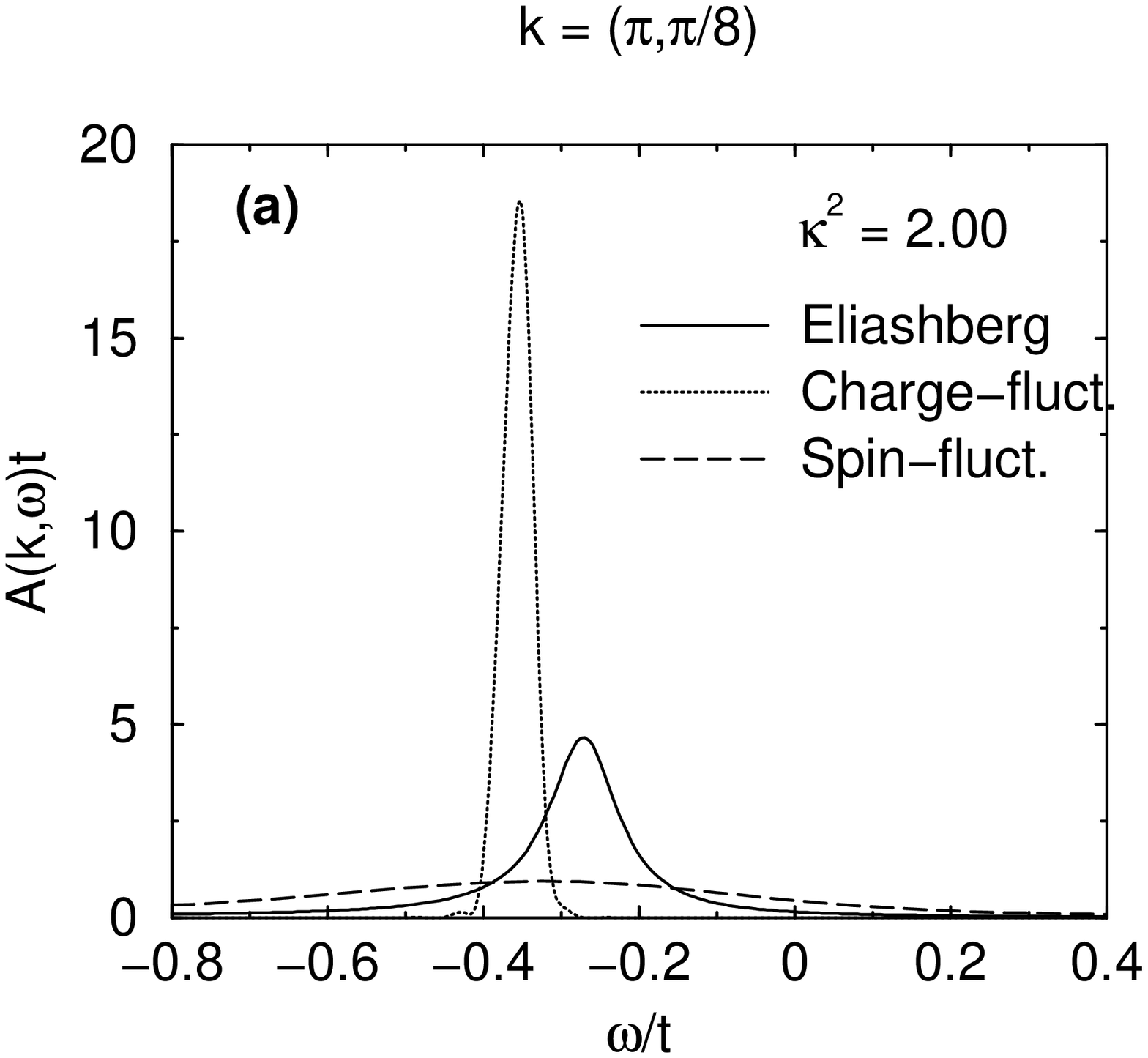}}
\centerline{\epsfysize=6.00in
\epsfbox{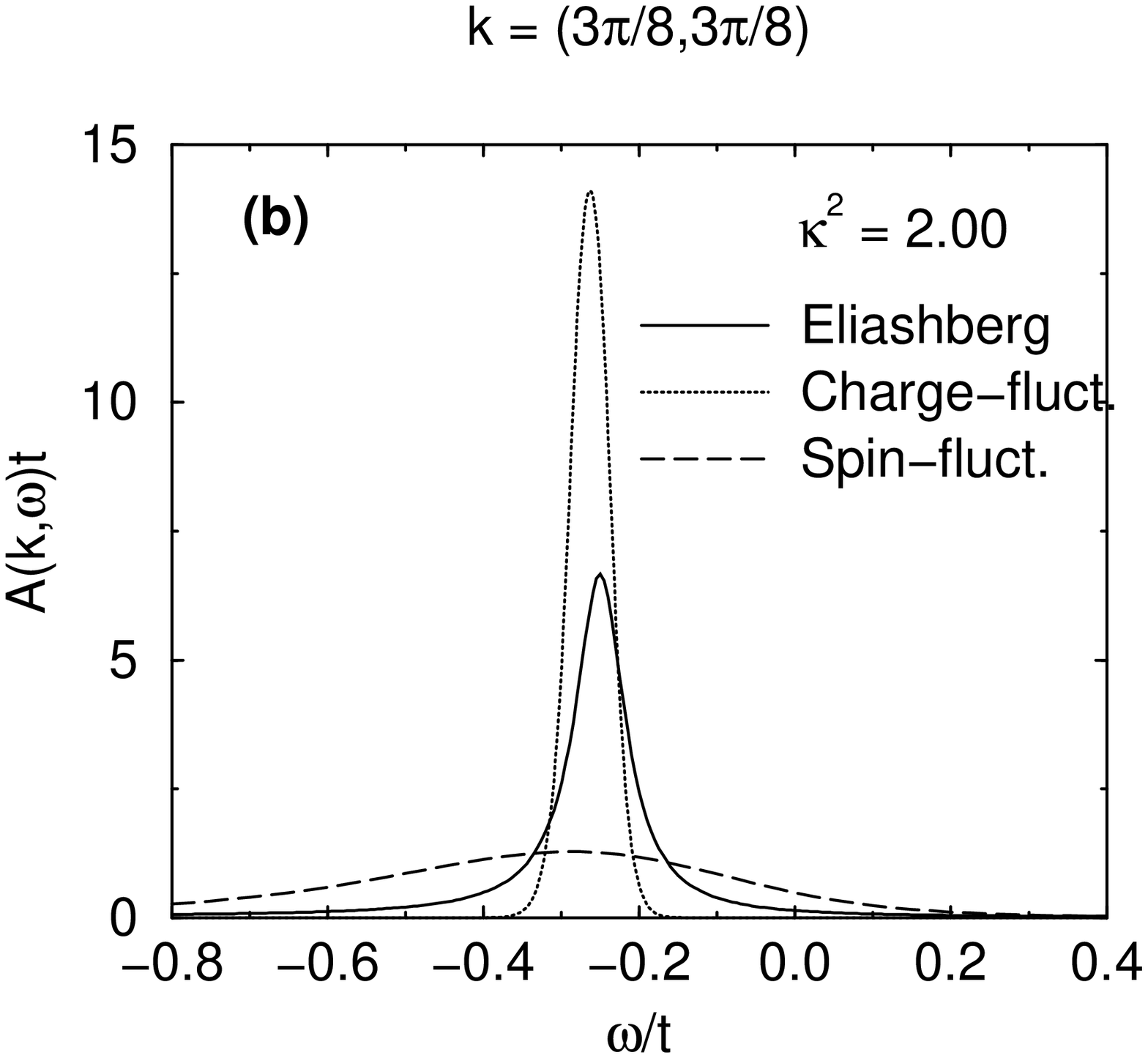}}
\label{fig5}
\caption{Comparison of the one-loop self-consistent versus the 
non-perturbative calculation of charge- and spin-fluctuation
exchanges. (a) shows the spectral function $A({\bf k},\omega)$ 
for a wavevector close to the Van-Hove singularity and (b) 
shows $A({\bf k},\omega)$ for a wavevector along the diagonal 
of the Brillouin zone.}
\end{figure}

\begin{figure}
\centerline{\epsfysize=6.00in
\epsfbox{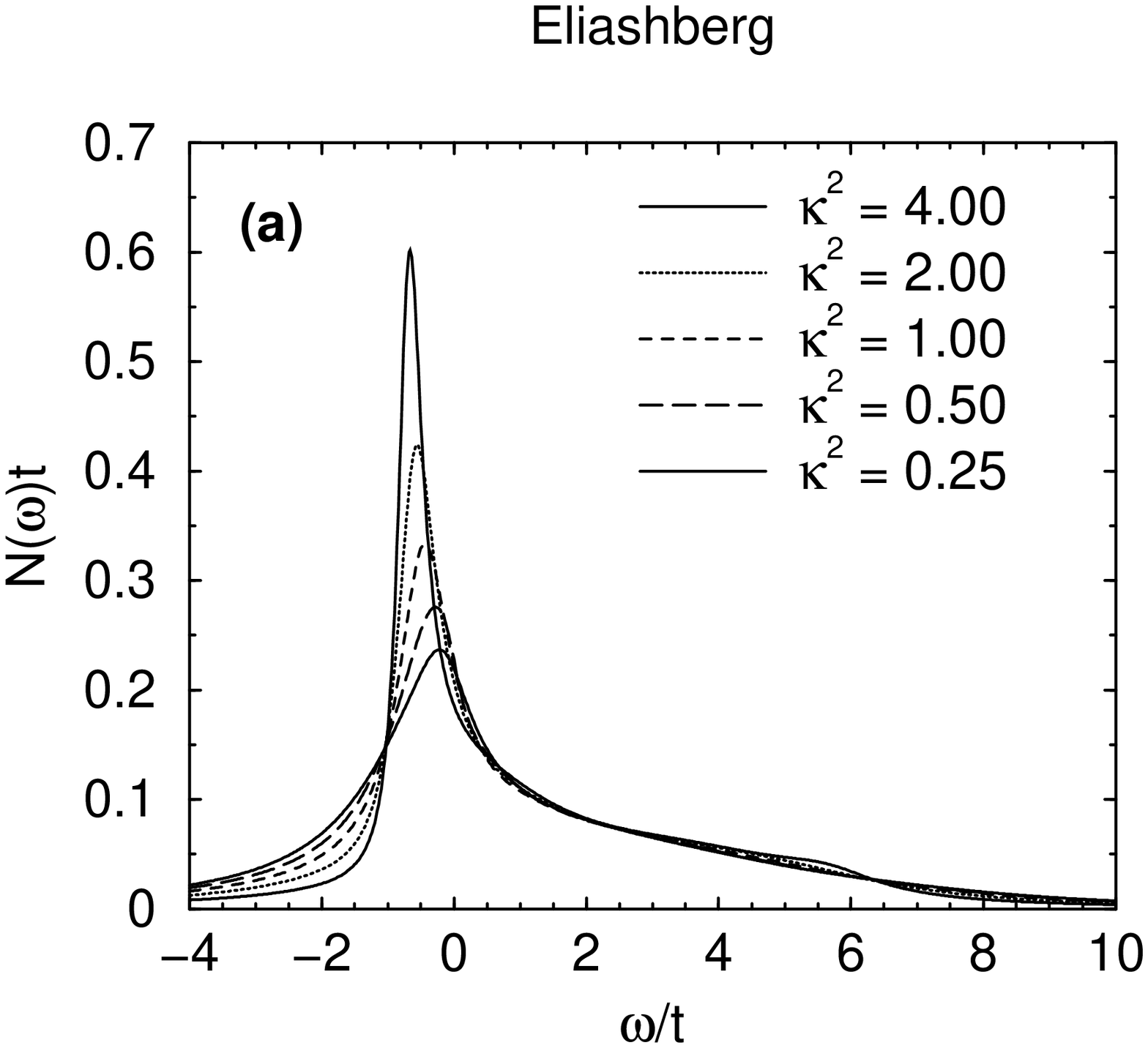}}
\centerline{\epsfysize=6.00in
\epsfbox{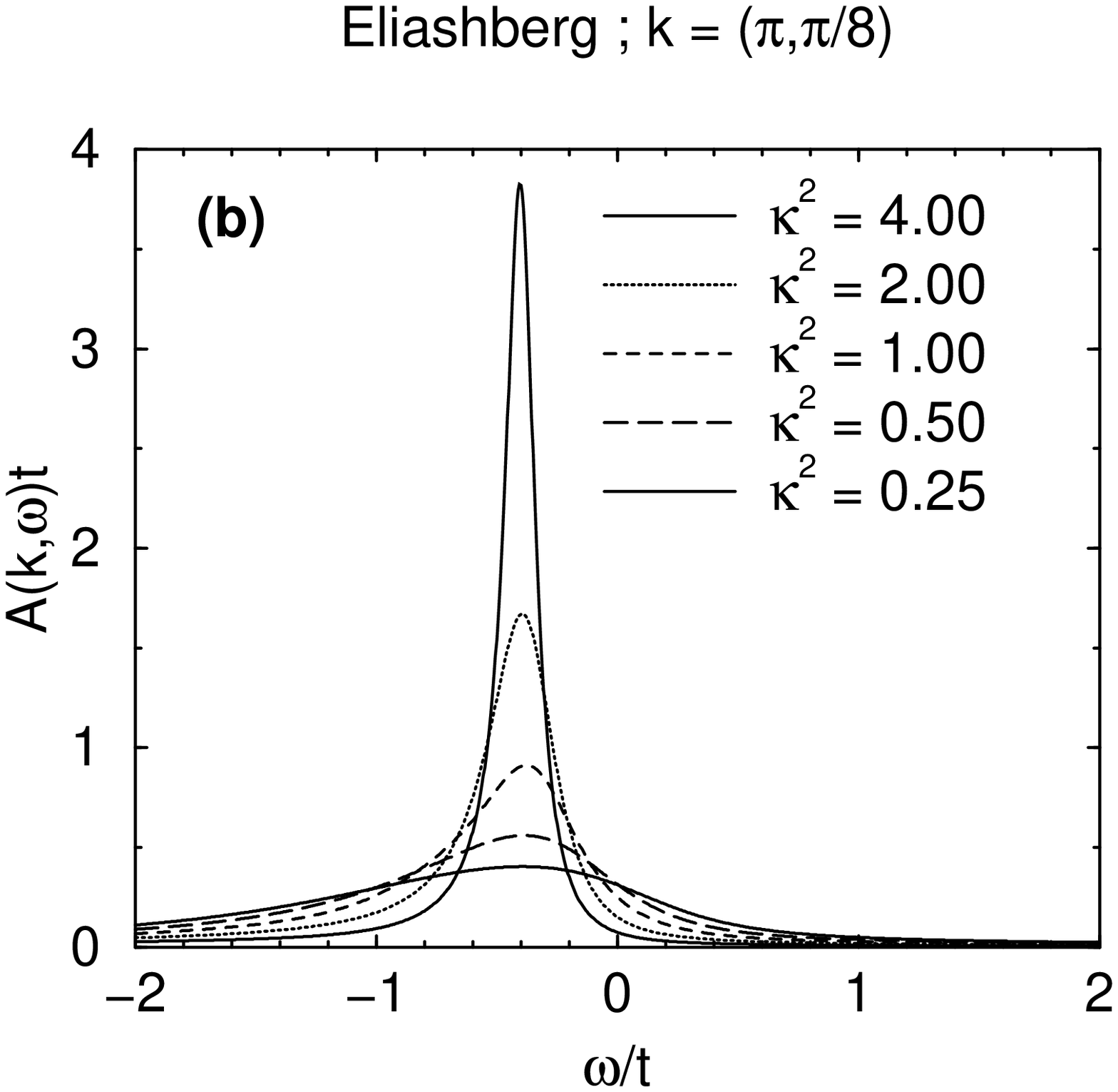}}
\centerline{\epsfysize=6.00in
\epsfbox{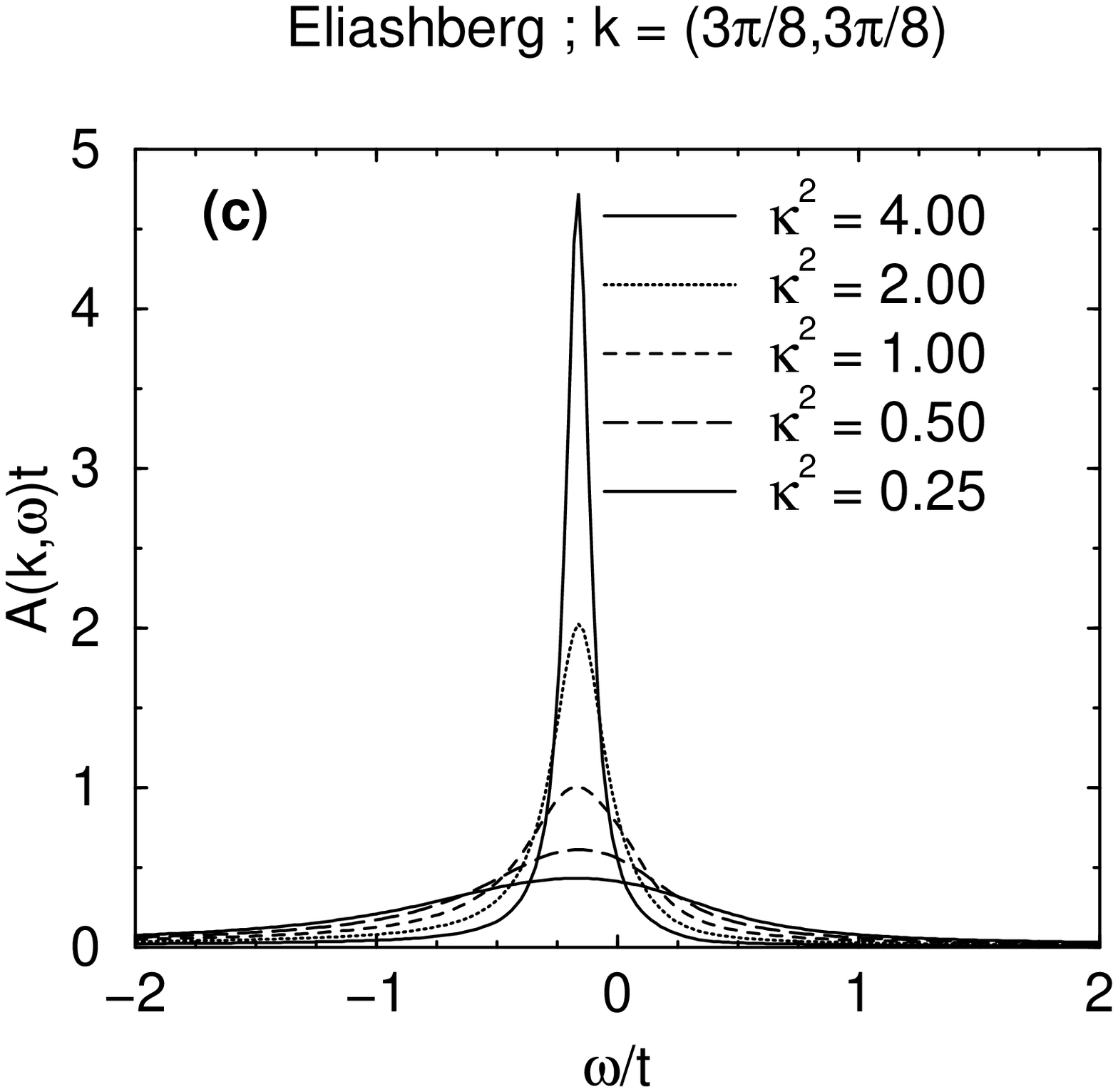}}
\label{fig6}
\caption{The one-loop approximation for the quasiparticle
properties for ferromagnetic spin-fluctuations.
The tunneling density of states $N(\omega)$ is shown in (a)
while (b) and (c) show the quasiparticle spectral function 
$A({\bf k},\omega)$ for momenta just below the Fermi level. 
(b) shows $A({\bf k},\omega)$ for a wavevector close to 
the Van Hove singularity and (c) shows $A({\bf k},\omega)$ 
for a wavevector along the diagonal of the Brillouin zone.}
\end{figure}

\begin{figure}
\centerline{\epsfysize=6.00in
\epsfbox{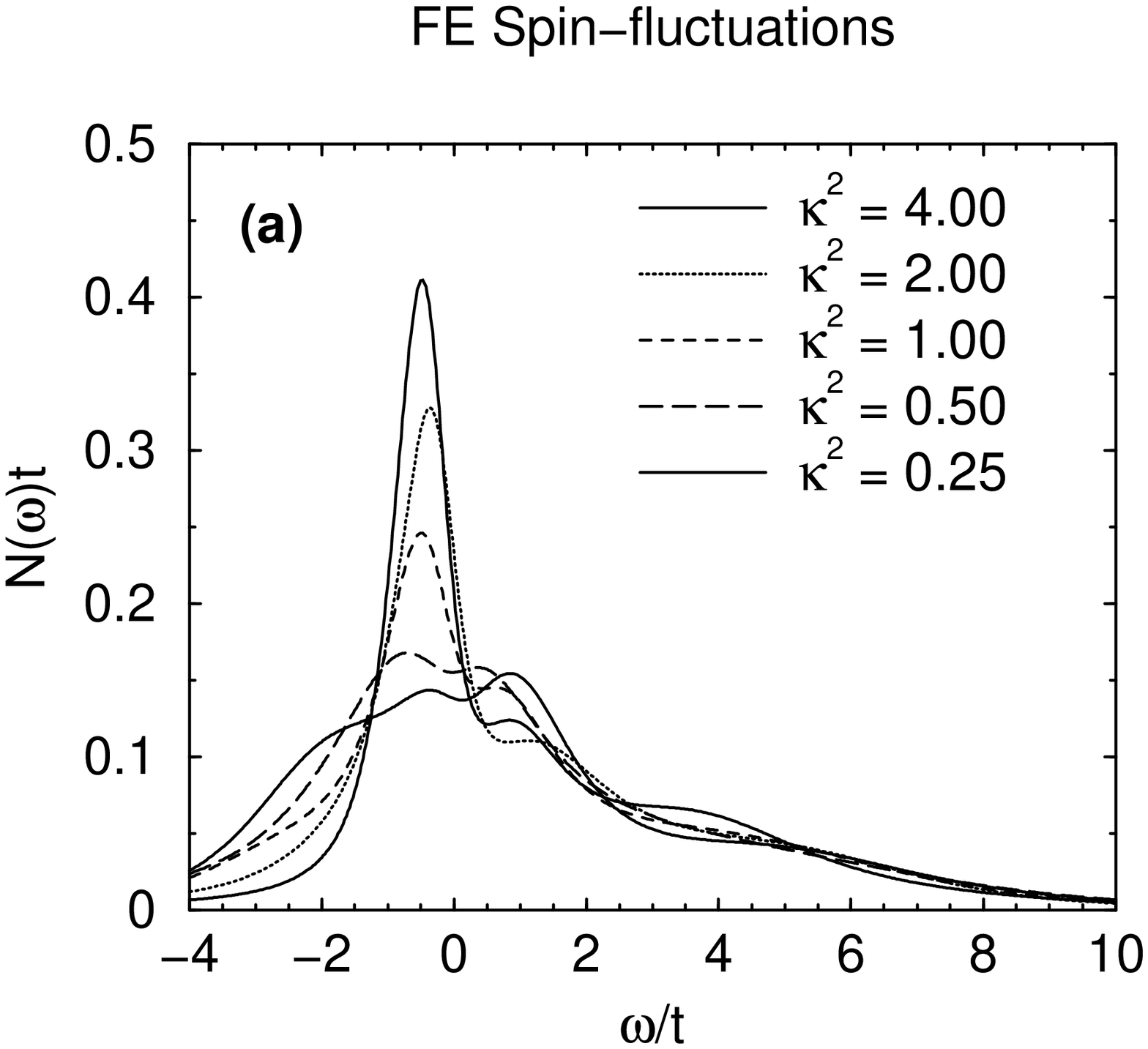}}
\centerline{\epsfysize=6.00in
\epsfbox{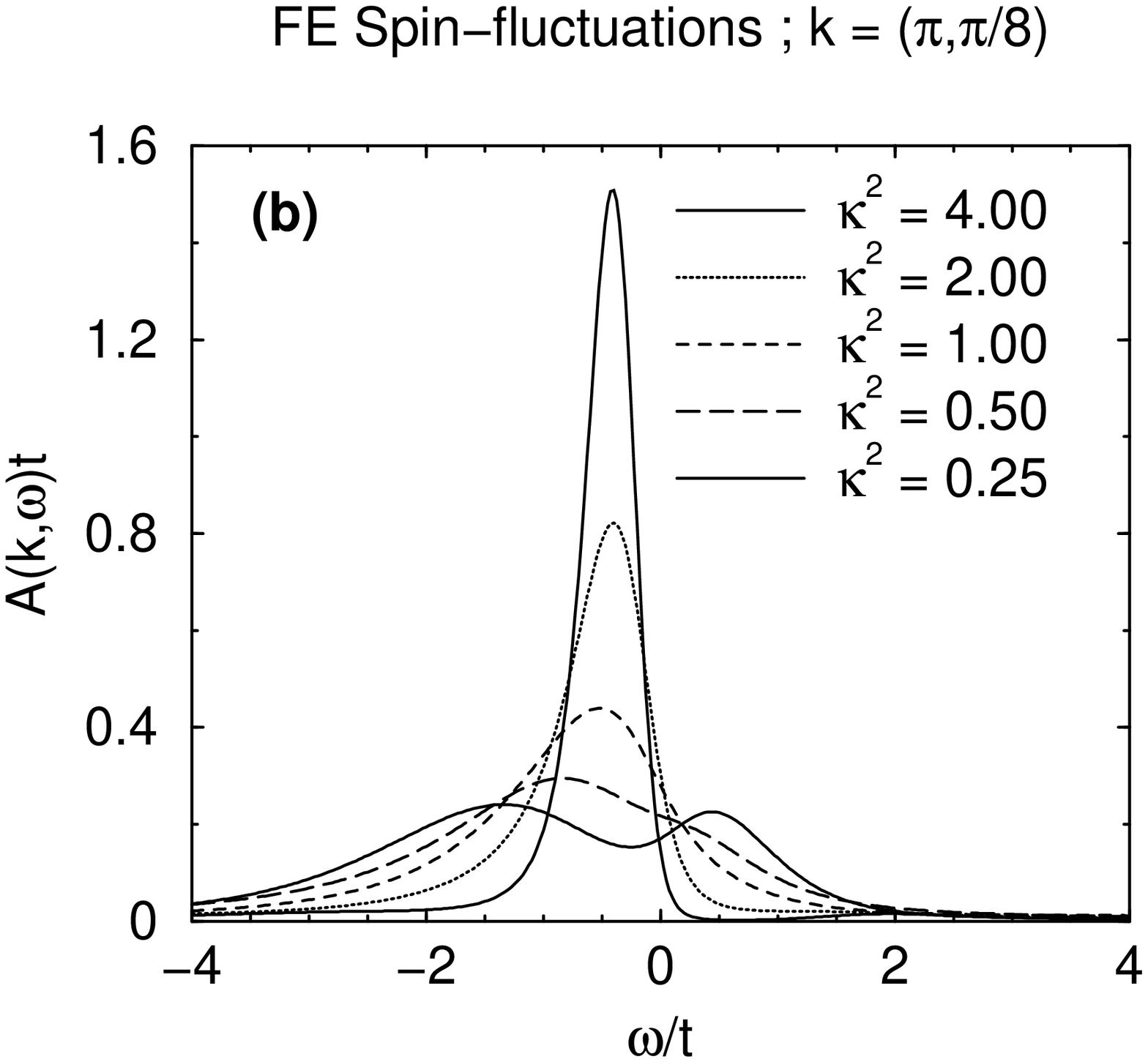}}
\centerline{\epsfysize=6.00in
\epsfbox{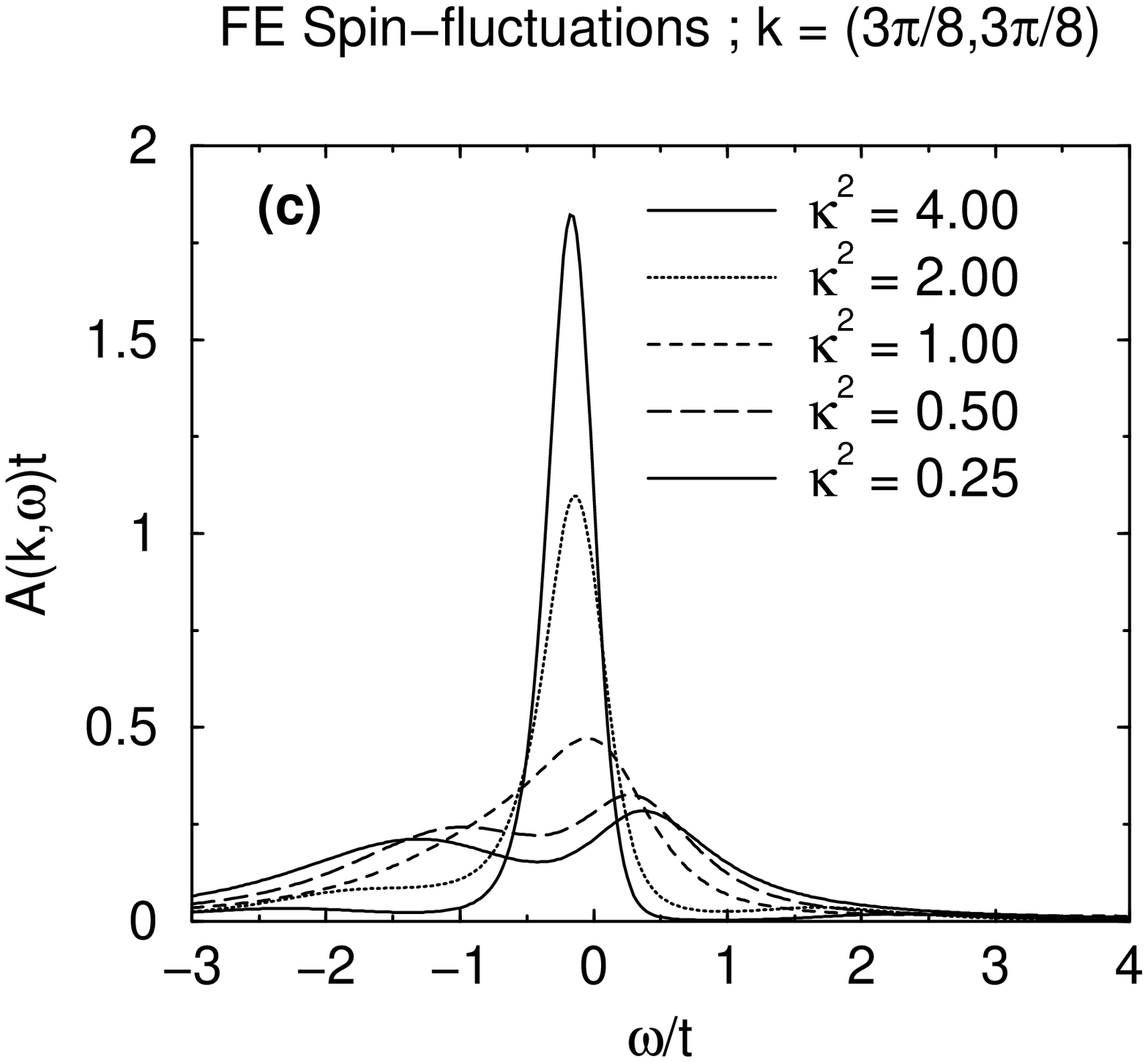}}
\label{fig7}
\caption{Quasiparticle properties for a coupling to an 
exchange vector dynamical molecular field ${\bf M}$
with ferromagnetic correlations. 
The tunneling density of states $N(\omega)$ is shown in 
(a) while (b) and (c) show the quasiparticle spectral 
function $A({\bf k},\omega)$ for momenta just below the 
Fermi level. (b) shows $A({\bf k},\omega)$ for a wavevector 
close to the Van Hove singularity and (c) shows $A({\bf k},\omega)$ 
for a wavevector along the diagonal of the Brillouin zone.}
\end{figure}

\begin{figure}
\centerline{\epsfysize=4.00in
\epsfbox{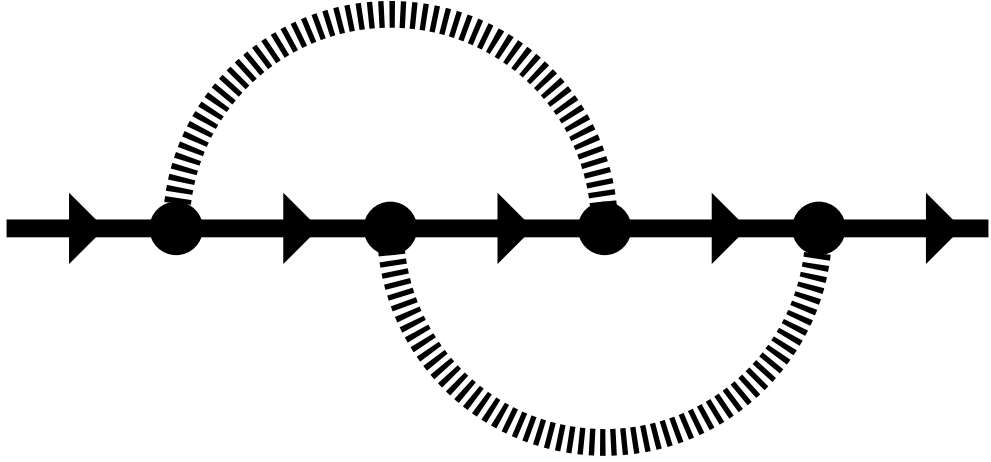}}
\label{fig8}
\caption{First order vertex correction to the one-loop
self-energy. The way spin is carried through the diagram
depends on the kind of molecular field. A Pauli matrix is 
associated with each vertex in the case of a coupling of 
quasiparticles to magnetic fluctuations. In the case of 
exchange of charge fluctuations, each vertex simply carries 
a unit matrix, namely the spin orientation is unchanged
at each vertex of the diagram.}
\end{figure}

\begin{figure}
\centerline{\epsfysize=6.00in
\epsfbox{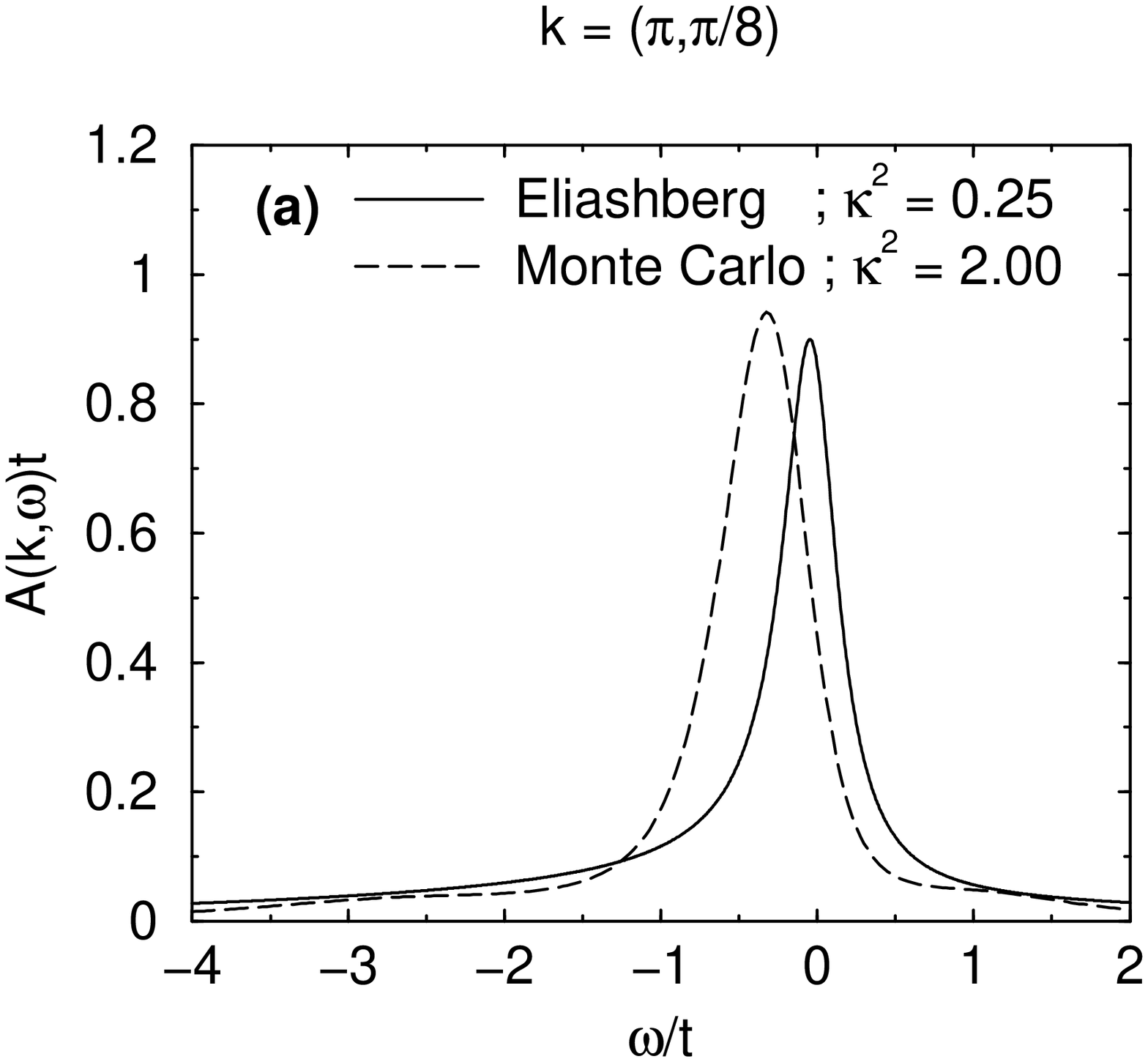}}
\centerline{\epsfysize=6.00in
\epsfbox{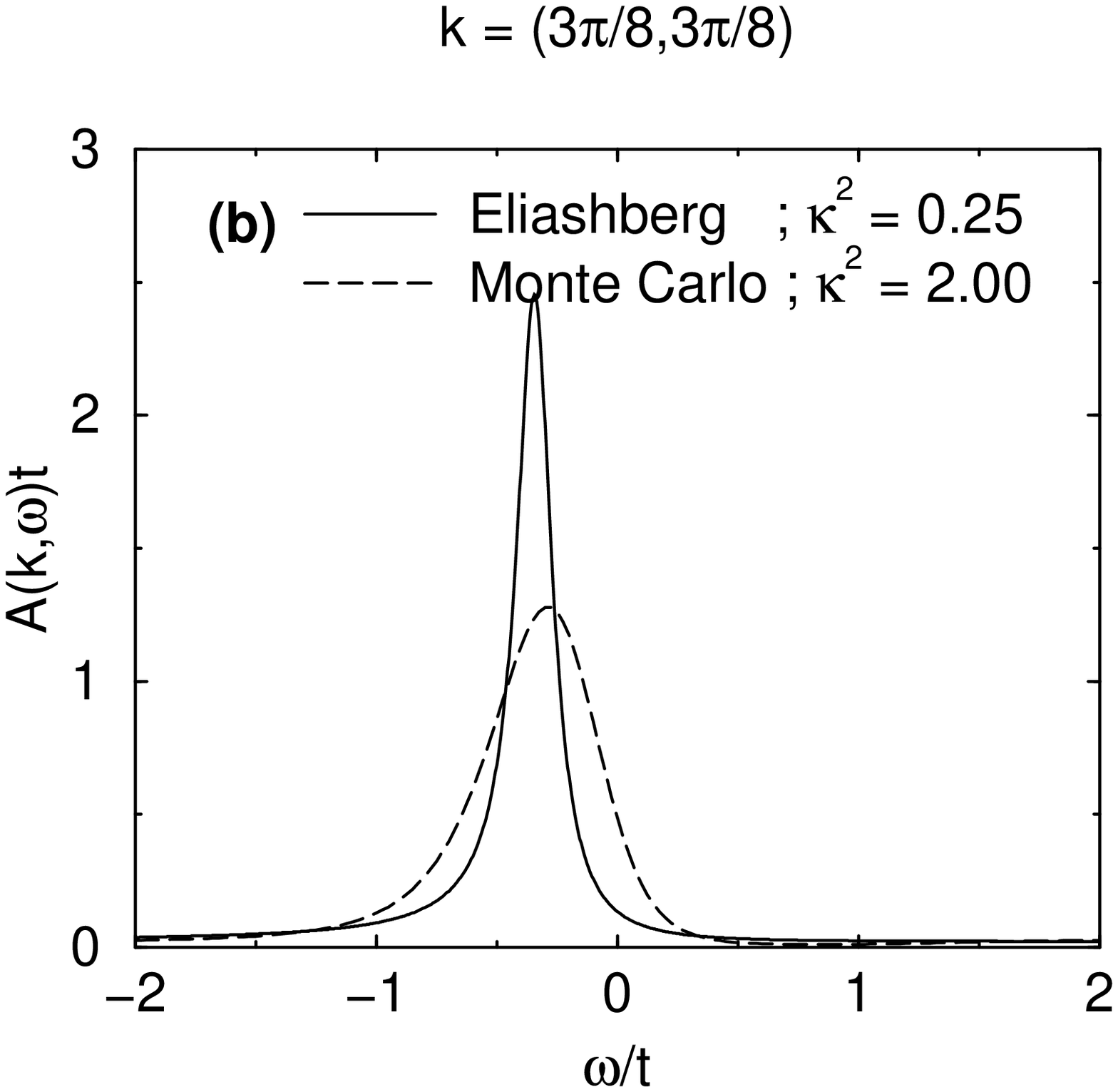}}
\label{fig9}
\caption{Comparison of the single antiferromagnetic
spin-fluctuation exchange approximation for the 
quasiparticle spectral function with a renormalized
correlation wavevector $\kappa^2_{eff} = 0.35$ and
the non-perturbative result with $\kappa^2 = 2$.
In both cases the dimensionless coupling constant
$g^2\chi_0/t = 2$.
(a) and (b) show the quasiparticle spectral function 
$A({\bf k},\omega)$ for momenta just below the Fermi level. 
(a) shows $A({\bf k},\omega)$ for a wavevector close to 
the Van Hove singularity and (b) shows $A({\bf k},\omega)$ 
for a wavevector along the diagonal of the Brillouin zone.}
\end{figure}

\begin{figure}
\centerline{\epsfysize=6.00in
\epsfbox{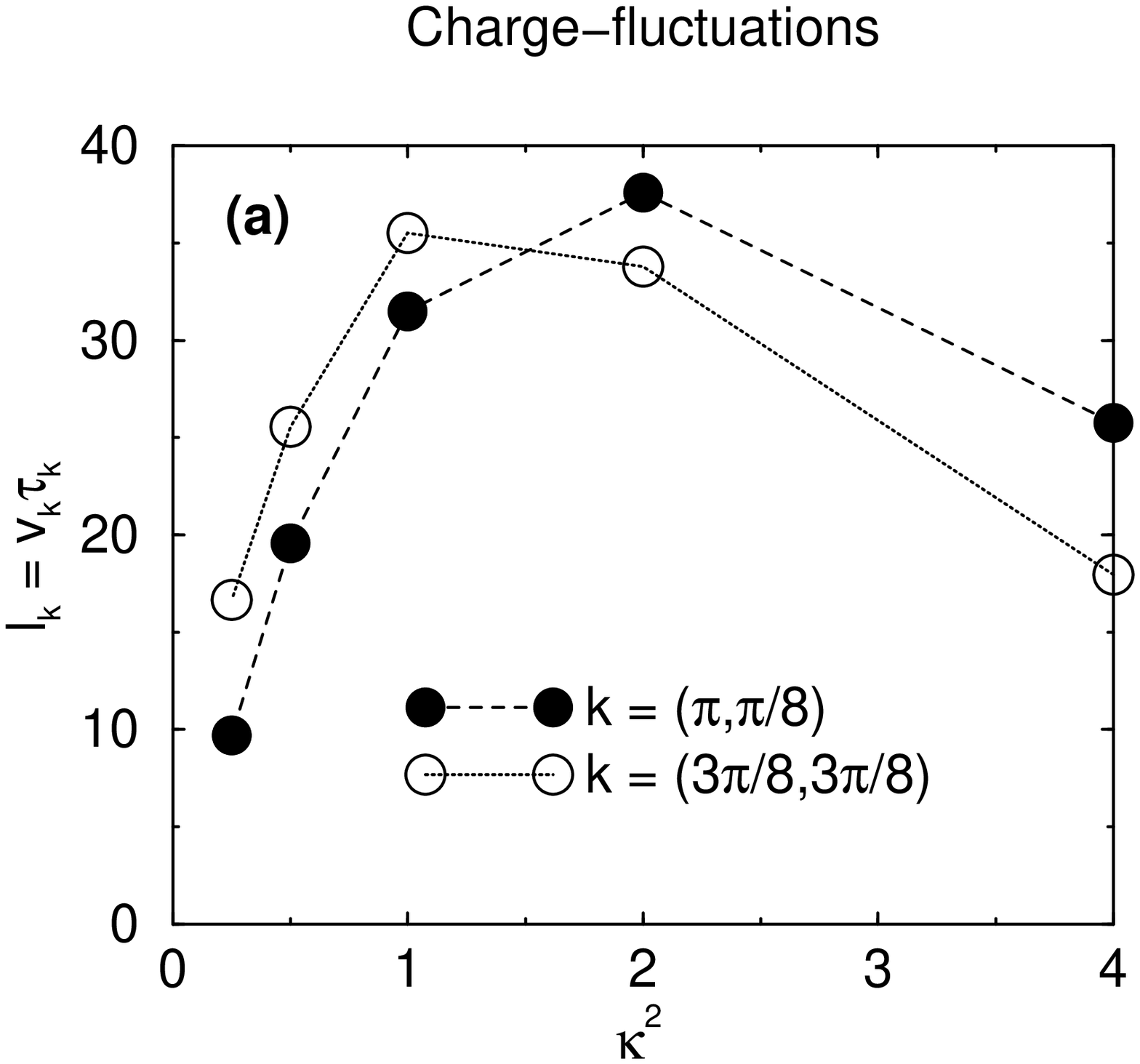}}
\centerline{\epsfysize=6.00in
\epsfbox{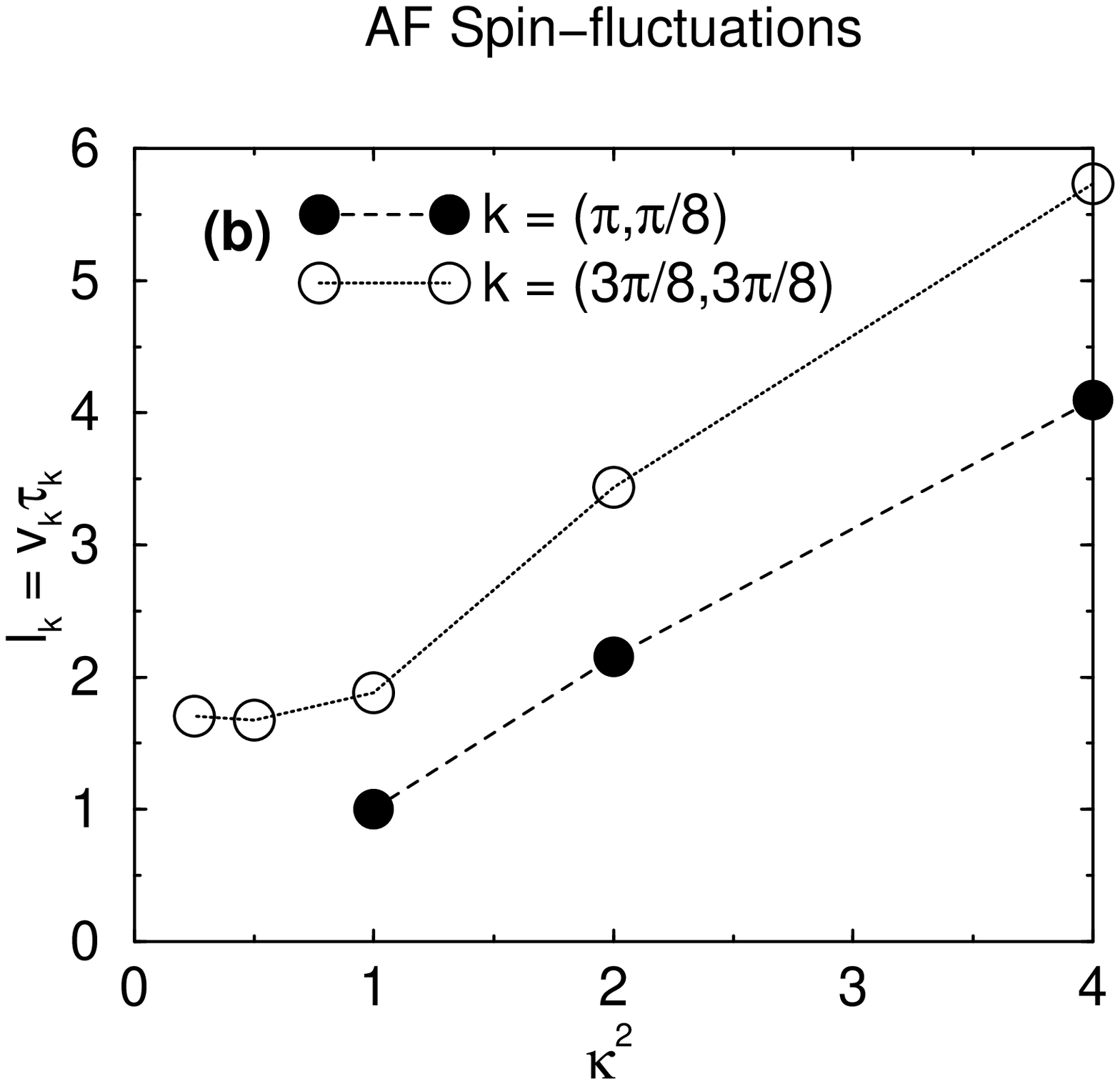}}
\centerline{\epsfysize=6.00in
\epsfbox{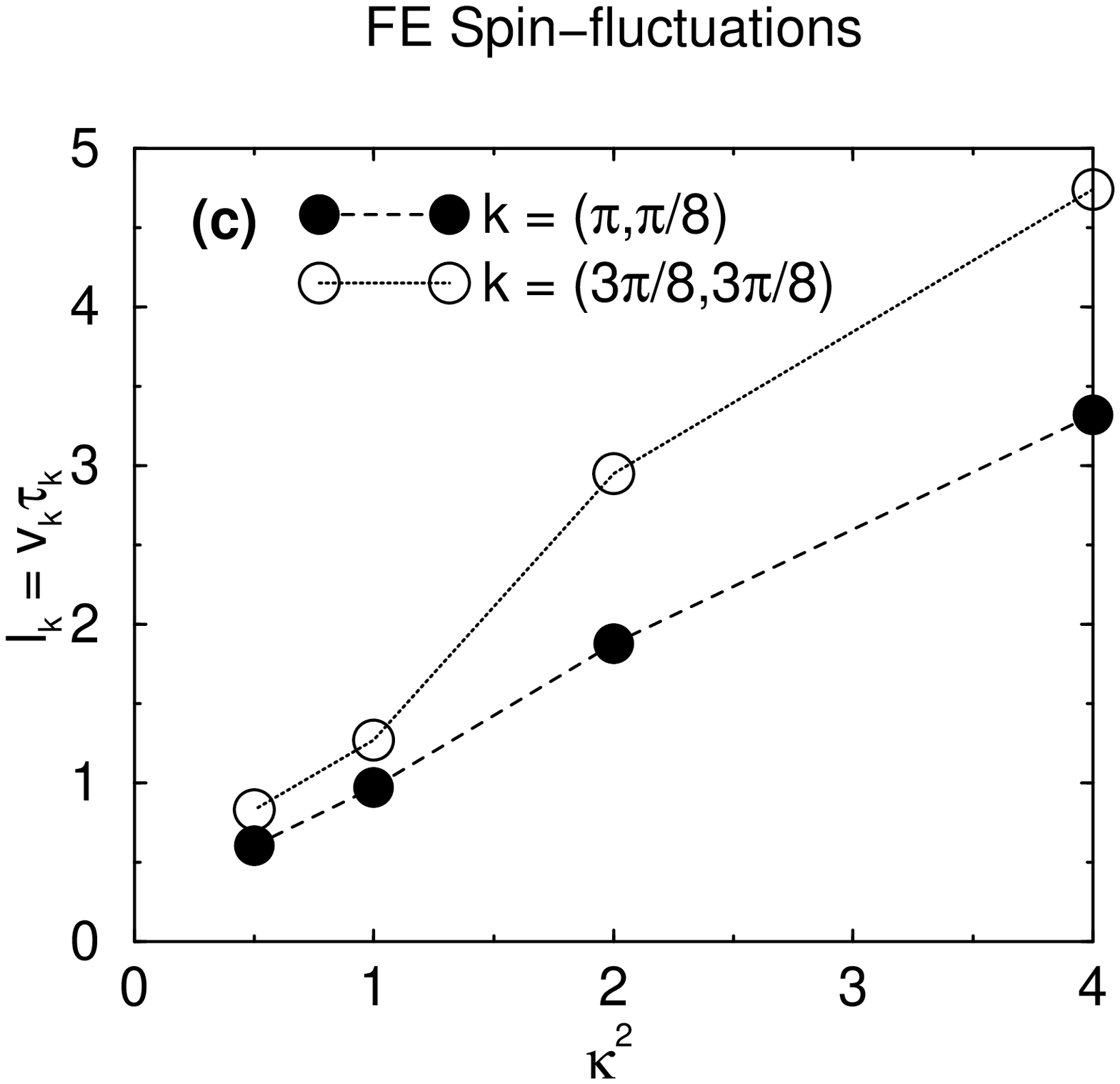}}
\label{fig10}
\caption{Approximate quasiparticle mean-free paths
$l_{\bf k}$ for ${\bf k} = (\pi,\pi/8)$ and 
${\bf k} = (3\pi/8,3\pi/8)$ obtained from Lorentzian 
fits to the numerical results for the spectral function. 
(a) shows $l_{\bf k}$ for coupling to charge-fluctuations, 
(b) to antiferromagnetic spin-fluctuations and (c)
to ferromagnetic spin-fluctuations.}
\end{figure}

\end{document}